\documentclass[11pt,a4paper]{article}
\usepackage{graphicx,subfig}
\pdfoutput=1
\usepackage{jheppub}
\usepackage{bm}
\usepackage{color}
\usepackage{fancyhdr}
\usepackage[yyyymmdd,hhmmss]{datetime}

\def\NeqFour{{\cal N} = 4}
\def\NeqFive{{\cal N} = 5}
\def\NeqEight{{\cal N} = 8}

\def\eqn#1{Eq.~(\ref{#1})}

\def\eqns#1#2{Eq.~(\ref{#1}) and~(\ref{#2})}
\def\fig#1{Fig.~{\ref{#1}}}

\def\sect#1{Section~{\ref{#1}}}
\def\tree{{\rm tree}}
\def\oneloop{{\rm one\hbox{-}loop}}
\def\twoloop{{\rm two\hbox{-}loop}}
\def\pol{\varepsilon}
\def\eps{\epsilon}

\title{Manifesting enhanced cancellations in supergravity: integrands versus integrals}

\author{Zvi Bern, Michael Enciso, Julio Parra-Martinez, Mao Zeng}
\affiliation{Mani L. Bhaumik Institute for Theoretical Physics,
Department of Physics and Astronomy,\\
University of California at Los Angeles, Los Angeles, CA 90095, USA}
\emailAdd{bern@physics.ucla.edu, menciso@physics.ucla.edu, jparra@physics.ucla.edu, zengmao@physics.ucla.edu}

\abstract{Examples of `enhanced ultraviolet cancellations' with no
  known standard-symmetry explanation have been found in a variety of
  supergravity theories.  By examining one- and two-loop examples in
  four- and five-dimensional half-maximal supergravity, we argue that
  enhanced cancellations in general cannot be exhibited prior to
  integration.  In light of this, we explore reorganizations of
  integrands into parts that are manifestly finite and parts that have
  poor power counting but integrate to zero due to integral
  identities.  At two loops we find that in the large loop-momentum
  limit the required integral identities follow from Lorentz and SL(2)
  relabeling symmetry.  We carry out a nontrivial check at four loops showing
  that the identities generated in this way are a complete set.  We
  propose that at $L$ loops the combination of Lorentz and SL($L$) symmetry
  is sufficient for displaying enhanced cancellations when they
  happen, whenever the theory is known to be ultraviolet finite up to
  $(L-1)$ loops.}

\keywords{scattering amplitudes, supergravity, ultraviolet, counterterms,
  loop integrals}

\begin{document}
\maketitle

\hfuzz=15pt
\section{Introduction}
\label{sec:intro}

The study of ultraviolet properties of four-dimensional gravity
theories has a long history, starting from the seminal work of
't~Hooft and Veltman~\cite{tHooftVeltman}. Despite this we do not know
the answer to the basic question of at which loop order various gravity
theories actually diverge.  In addition, when divergences occur in graviton
amplitudes we now know that they have unusual properties, including
dependence on evanescent effects~\cite{TwoLoopEvanescent} and
suspected links to anomalies~\cite{CarrascoAnomaly, N4GravFourLoops}.
Even more interesting are indications in certain supergravity theories
that the loop order where the first divergence occurs is higher than
previous expectations~\cite{N4GravThreeLoops, N5GravFourLoops,
  TwoLoopHalfMaxD5}.  This renews the possibility that certain
theories, such as $\NeqEight$ supergravity, are ultraviolet finite at
any order in perturbation theory. No known symmetry is powerful enough
to render a four-dimensional quantum gravity theory ultraviolet
finite, so if this were true it would be extraordinary.

Certain cancellations in gravity theories are different from those in
supersymmetric gauge theories in that they cannot be made manifest for
ordinary local representations. When such cancellations happen they
are dubbed `enhanced cancellations'~\cite{N5GravFourLoops}.  In simple
cases, these enhanced cancellations can be understood through
conventional means by constraining the set of available counterterms
from symmetry considerations.  For example, at one loop, a well-known
counterterm argument~\cite{tHooftVeltman} explains that the $n$
graviton amplitudes are finite even though the diagrams scale poorly
in the ultraviolet.  On the other hand, recent examples of enhanced
cancellations have as yet no standard symmetry explanation, despite
attempts~\cite{VanishingVolume,KellyAttempt,HalfMaxMatter} and insight from string
theory~\cite{PierreN4}.  These examples include $\NeqFive$
supergravity at four loops in $D=4$~\cite{N5GravFourLoops}, $\NeqFour$
supergravity at three loops in $D=4$~\cite{N4GravThreeLoops}, and
half-maximal supergravity at two loops in
$D=5$~\cite{TwoLoopHalfMaxD5}.  In the relatively simple case of
half-maximal supergravity at two loops the cancellations have been
understood using the double-copy structure that allows the amplitudes
to be built from gauge-theory ones~\cite{TwoLoopHalfMaxD5}.
Unfortunately, it is not clear how to generalize this understanding to
higher loops.

In light of the difficulties in trying to develop a comprehensive
explanation for enhanced cancellations, we should consider alternative
approaches.  For instance one could try to mimic diagram-based proofs
of finiteness that were successfully carried out for $\NeqFour$
super-Yang--Mills theory (see for example
Refs.~\cite{Finiteness,UVProofs}).  These were achieved by finding
representations of the integrand where every term is ultraviolet
finite by power counting.  However, enhanced cancellations are
different: By definition they cannot be made manifest diagram by
diagram at the integrand level, using only standard Feynman
propagators.  But one can still wonder if some kind of integrand-level
reorganization could be found that makes large loop-momentum
cancellations manifest or at least clarifies how the cancellations
occur.

An obstruction to pursuing these ideas is that we lack a good definition of global
variables for all diagrams of a multiloop amplitude including
nonplanar diagrams.
One way to approach this difficulty is to use unitarity cuts. At one
loop, a systematic program was successfully followed for all one-loop
(super)gravity amplitudes in Ref.~\cite{UnexpectedCancellations} using
a formalism~\cite{Forde} based on generalized
unitarity~\cite{GeneralizedUnitarity}.  This was used to demonstrate
the existence of nontrivial cancellations between diagrams as the
number of external legs increases. However, a general extension of the
one-loop analysis to higher loops remains a challenge.

In this paper instead of attempting a general argument we turn to
specific examples in half-maximal supergravity, which we study in some
detail.  We construct the examples using the Bern--Carrasco--Johansson
(BCJ) double-copy construction of gravity loop integrands in terms of
gauge-theory ones~\cite{BCJ,BCJLoop}.  These examples are based
on the one- and two-loop $\NeqFour$ supergravity amplitudes previously
obtained in Refs.~\cite{DunbarN4,BernOneLoopN4,DixonTwoLoopN4}.

We first show that at one loop it is not possible to construct
integrands where cancellations are manifest in general dimensions. In particular, we identify cancellations in $D=4$ that require integration identities.  At two loops we use unitarity cuts to argue that cancellations cannot be made manifest at the integrand level. To further investigate this case, we use integration-by-parts (IBP) technology \cite{IBP, KosowerIBP, ItaIBP, LarsenZhang} to reorganize the integrand into pieces that are finite by power
counting and pieces that are divergent by power counting, yet
integrate to zero.  Although this re-arrangement of the complete
integrand is successful, it requires detailed knowledge of the
specific integrals and their relations, making it difficult to
generalize to higher loops.

To deal with this, we then turn to a simpler approach by giving up on
trying to make the full integrand display enhanced ultraviolet
cancellations. Instead we series expand in large loop momenta in order
to focus on the ultraviolet behavior.  We show that at least in the
two loop examples we study the integral identities necessary for
exposing the enhanced cancellations follow from only Lorentz and SL(2)
relabeling invariance.  These ideas continue to higher loops, and as a nontrivial confirmation we found that
these principles generate all required integral identities for
exposing the ultraviolet behavior of maximal supergravity at four
loops in the critical dimension where the divergences first
occur~\cite{Simplifying}.  Based on these results, we conjecture that
at $L$ loops the IBP identities generated by Lorentz and SL($L$)
relabeling symmetry are sufficient for revealing the enhanced
cancellations, when they exist. The principles are generic
and present in all amplitudes in the large loop-momentum limit.

This paper is organized as follows.  In \sect{sec:example}, we present
one- and two-loop examples showing the lack of integrand-level
cancellations.  In \sect{sec:rearranging} we outline how one can
arrange complete integrands so that they are manifestly finite by
power counting up to terms that integrate to zero.  In \sect{sec:ibp}
we then analyze the large loop-momentum limit, bringing us to a
conjecture on symmetries of the integrals responsible for making
enhanced cancellations visible.  We give our conclusions
in \sect{sec:conc}.  We also include an appendix on subtleties
regarding boundary terms in integration-by-parts identities.


\section{Absence of enhanced cancellations in the integrand}
\label{sec:example}

Enhanced cancellations are a recently identified type of ultraviolet
cancellation that can occur in gravity
theories~\cite{N4GravThreeLoops, N5GravFourLoops,
  TwoLoopHalfMaxD5}. These cancellations are defined as follows: Start
with an amplitude organized in terms of diagrams whose denominators
are only the usual Feynman propagators $i/(p^2 + i \eps)$.  Suppose
this amplitude is ultraviolet finite, yet there are terms that are
divergent by power counting and cannot be re-assigned to other
diagrams without introducing additional spurious denominators in other
diagrams.  This implies nontrivial cancellations that cannot be
manifest in the integrand of each diagram.  We would then say there is
an enhanced cancellation.

This notion is distinct from the question of whether it is possible to
exhibit the cancellations at the integrand level; one
might imagine that with careful choices of loop variables in each
diagram, one might be able to align the loop momenta in just the right
way so that poor behavior cancels algebraically between diagrams prior
to integration.  Here we show that this does not happen.

We present examples of enhanced cancellations to
illustrate that it is only after integration that divergences cancel.
We focus on the relatively simple cases of 16-supercharge half-maximal
supergravity at one and two loops in $D=4$ and $D=5$. In $D=4$ this
theory is just $\NeqFour$ supergravity~\cite{N4Sugra}.  Even though
the one-loop $D=4$ cancellation is a well-known consequence of
supersymmetry~\cite{OneLoopSugraDiv}, it provides a relatively simple
concrete example of cancellations that do not arise at the integrand
level, but can be exposed using Lorentz invariance.
We then turn to the more interesting case of two-loop half-maximal
supergravity in $D=5$. In this case no known standard-symmetry
argument invalidates the potential $R^4$
divergence~\cite{VanishingVolume,KellyAttempt,HalfMaxMatter}.

In order to construct the integrands we use the BCJ double-copy
construction~\cite{BCJ,BCJLoop}, which we review briefly.  The
double-copy construction is useful because it directly gives us
gravity loop integrands from corresponding gauge-theory ones.  In this
construction, one of the two gauge-theory amplitudes is first
reorganized into diagrams with only cubic vertices,
\begin{equation}
\mathcal{A}_m^{L\hbox{-}\rm loop} = i^L g^{m-2+2L} \sum_{{\cal S}_m} \sum_j
\int \prod_{l=1}^L \frac{d^D p_l}{(2\pi)^D} \frac{1}{S_j}
\frac{c_j n_j}{\prod_{\alpha_j} D_{\alpha_j}}\,,
\label{GaugeCubic}
\end{equation}
where the $D_{\alpha_j}$ are the propagators of the $j^{\rm th}$ diagram,
$L$ is the number of loops, $m$ is the number of external legs and
$g$ is the gauge coupling. The first sum runs over the $m!$ permutations
of external legs, denoted by ${\cal S}_m$, while the second sum over
$j$ runs over the distinct cubic graphs.  The product in the
denominator runs over all Feynman propagators.  The symmetry factor
$S_n$ accounts for any overcounts and internal automorphisms.
The $c_j$ are the color factors associated with the diagrams and
the $n_j$ are kinematic numerators.

The double-copy construction relies on BCJ duality~\cite{BCJ,BCJLoop} where
triplets of diagram numerators satisfy equations in one-to-one
correspondence with the Jacobi identities of the color factors of each
diagram,
\begin{equation}
c_i + c_j + c_k = 0 \, \Rightarrow \, n_i + n_j + n_k = 0\,.
\label{BCJDuality}
\end{equation}
The indices $i, j, k$ label the diagram to which the color factors and
numerators belong.  If the diagram numerators satisfy the same
algebraic properties as the color factors, we can obtain corresponding
gravity amplitudes simply replacing the color factors of a second
gauge theory with numerator factors where the duality holds:
\begin{equation}
c_i \rightarrow n_i \,.
\label{ColorSubstitute}
\end{equation}
The gauge-theory coupling constant is also replaced by the
gravitational one: $g \rightarrow (\kappa/2)$.  In this construction
the duality (\ref{BCJDuality}) needs to be manifest in only one of the
two gauge theories~\cite{BCJLoop,Square}. This construction also
extends to cases where the gauge theory
includes fundamental-representation matter
particles~\cite{HenrikFundamental}.


\subsection{One-loop example}

\begin{figure}[tb]
  \centering
  \includegraphics[width=0.5\textwidth]{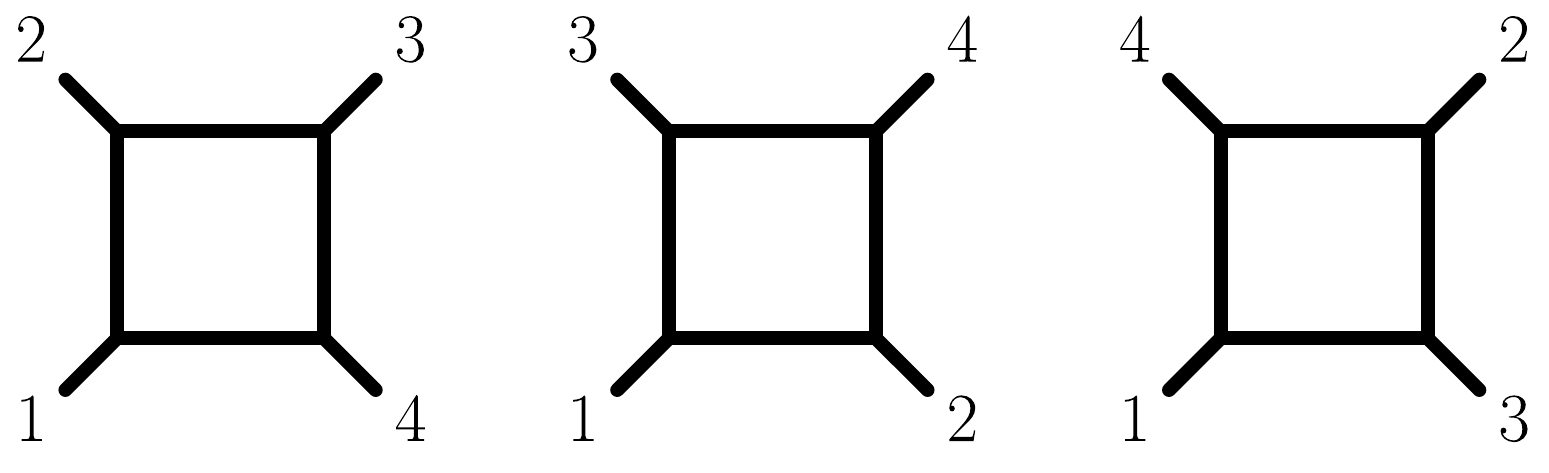}
  \caption{The three box diagrams contributing to the one-loop
    four-point amplitude of maximal $\NeqFour$ super-Yang--Mills theory and
    half-maximal supergravity.}
  \label{fig:box}
\end{figure}

We start with the one-loop amplitude of pure half-maximal $\NeqFour$
supergravity in four dimensions~\cite{N4Sugra}. This amplitude is well
studied and has been computed in Refs.~\cite{DunbarN4,BernOneLoopN4}.
The double-copy construction of this amplitude is particularly
simple. We start from the corresponding $\NeqFour$ super-Yang--Mills
and pure Yang--Mills amplitudes.

The one-loop four-point $\NeqFour$ super-Yang--Mills amplitude was
first obtained from the low-energy limit of a Type~I superstring
amplitude~\cite{GSB}.  This amplitude is particularly simple and the
only nonzero kinematic numerators are those of the box diagrams
in \fig{fig:box},
\begin{equation}
n^{\rm box}_{\NeqFour} = s t A^{\rm tree}_{\NeqFour}(1,2,3,4) \,,
\label{N4YMNumerator}
\end{equation}
where $s = (k_1+k_2)^2$ and $t=(k_2+k_3)^2$ are the usual Mandelstam
invariants and $A^{\rm tree}_{\NeqFour}(1,2,3,4)$ is the color-ordered
tree superamplitude. The combination $s t A^{\rm
tree}_{\NeqFour}(1,2,3,4)$ is crossing symmetric, so the three box diagrams
have identical numerators.  It is easy to check that this
representation of the amplitude satisfies the color-kinematics
duality (\ref{BCJDuality}).

Replacing the color factors in the pure Yang--Mills box contributions
given in Ref.~\cite{ColorKinOneTwoLoops} with the $\NeqFour$
super-Yang--Mills numerators (\ref{N4YMNumerator}), we obtain the
$\NeqFour$ supergravity amplitude as a sum over box diagrams,
\begin{equation}
{\cal M}^{\oneloop}_{\NeqFour} = - \biggl(\frac{\kappa}{2}\biggr)^4
st A^\tree_{{\NeqFour}}(1,2,3,4) \Bigl(I_{1234}[n_{1234,p}] + I_{1324}[n_{1324,p}]
          + I_{1423}[n_{1423,p}]\Big) \,,
\end{equation}
where
\begin{equation}
I_{1234}[n_{1234,p}] \equiv
\int \frac{d^D p}{(2\pi)^D} \frac{n_{1234,p}}{p^2 (p-k_1)^2 (p-k_1-k_2)^2
      (p+k_4)^2} \,,
\end{equation}
is the first box integral in \fig{fig:box} and ${n_{1234,p}}$ is the
expression defined in Eq.~(3.5) of
Ref.~\cite{ColorKinOneTwoLoops}. The triangle and bubble contributions
from the pure Yang--Mills amplitude are simply set to zero because the corresponding
$\NeqFour$ SYM numerators vanish.  As dictated by the double-copy
construction, the supergravity states are given by the tensor product
of pure Yang--Mills gluon states with the states of $\NeqFour$
super-Yang--Mills theory.

The case of $D=4$ is an example of enhanced cancellations because the
three box diagrams are each logarithmically divergent, yet the sum
over diagrams is finite.  We can see this by finding power-counting
divergent terms in each diagram that cannot be moved to other diagrams
without introducing nonlocalities in the diagram numerators.  An
example is the term,
\begin{equation}
n_{1234,p} \sim  p_{\mu_1}  p_{\mu_2}  p_{\mu_3} p_{\mu_4}
      \pol^{\mu_1}_1  \pol^{\mu_2}_2 \pol^{\mu_3}_3 \pol^{\mu_4}_4 + \cdots \,,
\label{NumeratorTerm}
\end{equation}
where $\pol^{\mu_i}_i$ is the gluon polarization of leg $i$ on the
pure Yang--Mills side of the double copy.

The cancellations between the diagrams are nontrivial. To see the
cancellation of the logarithmic divergences, we expand in large loop
momentum or equivalently small external momenta $k_i^{\mu}$. Because
the integrals are only logarithmically divergent in $D=4$,
this amounts to simply setting all $k_i^{\mu}$ to zero in the
integrand (keeping the overall prefactor fixed).  In this limit, the
propagator of each graph become identical, and the resulting graph
effectively becomes a scaleless vacuum integral.  Such scaleless
integrals vanish in dimensional regularization, but we can introduce a
mass for each propagator to separate out the infrared divergences
without affecting the ultraviolet divergence.  Starting with the pure
Yang--Mills numerators, keeping only the leading terms in all three
box diagrams results in an integrand proportional to
\begin{align}
\hskip -.3 cm
-i st A^\tree_{{\NeqFour}} (D_s-2) \nonumber
  & \frac{\pol^{\mu_1}_1\pol^{\mu_2}_2\pol^{\mu_3}_3\pol^{\mu_4}_4}
    {2 (p^2 - m^2)^4}
 \Bigl[(p^2)^2(\eta_{\mu_1\mu_4}\eta_{\mu_2\mu_3}+\eta_{\mu_1\mu_3}
  \eta_{\mu_2\mu_4}+\eta_{\mu_1\mu_2}\eta_{\mu_3\mu_4})\\
-&4 p^2 (\eta_{\mu_1\mu_2}p_{\mu_3}p_{\mu_4} + \eta_{\mu_1\mu_3}p_{\mu_2}p_{\mu_4}
        + \eta_{\mu_1\mu_4}p_{\mu_2}p_{\mu_3} + \eta_{\mu_2\mu_3}p_{\mu_1}p_{\mu_4} \nonumber \\
      &  + \eta_{\mu_2\mu_4}p_{\mu_1}p_{\mu_3} + \eta_{\mu_3\mu_4}p_{\mu_1}p_{\mu_2} )
       + 24\, p_{\mu_1}p_{\mu_2}p_{\mu_3}p_{\mu_4} \Bigr]\,,
\label{OneLoopIntegrand}
\end{align}
where $A^\tree_{{\NeqFour}} = A^\tree_{{\NeqFour}}(1,2,3,4)$ and $D_s$ is a
state-counting parameter coming from contractions
$\eta_{\mu}{}^\mu = D_s$. (In conventional dimensional
regularization~\cite{CollinsBook} $D_s = 4-2\epsilon$, but in other
schemes, such as the four-dimensional helicity scheme~\cite{FDH}, $D_s =
4$.)  In the expression above we see explicitly that the amplitude is
logarithmically divergent by power counting and that no purely
algebraic manipulations can expose the cancellation of the divergence.
What makes this case particularly simple is that in the large
loop-momentum limit all diagrams degenerate to a single vacuum
integral, avoiding loop-momentum labeling ambiguities in different
terms that plague higher loops.

This example provides a clear demonstration that even after summing
over diagrams, enhanced cancellations are not visible prior to using
properties of integrals.  To expose the ultraviolet cancellation we
use Lorentz invariance in the form of integration identities:
\begin{align}
\int d^D p \, \frac{p_\mu p_\nu}{(p^2 - m^2)^4} &= \int d^D p \,
\frac{1}{D} \frac{\eta_{\mu \nu} p^2}{(p^2-m^2)^{4}} \,, \\
\int d^D p \, \frac{p_\mu p_\nu p_\rho p_\sigma}{(p^2 -m^2)^4}
&= \int d^D p \, \frac{1}{D(D+2)}
 \frac{(\eta_{\mu \nu} \eta_{\rho \sigma} + \eta_{\mu \rho} \eta_{\nu \sigma} +
\eta_{\mu \sigma} \eta_{\rho \nu} ) (p^2)^2}
        {(p^2-m^2)^{4}} \,.
\end{align}
With these identities, we find that the integral
of \eqn{OneLoopIntegrand} is equal to the integral of
\begin{align}
\nonumber -i st A_{{\NeqFour}}^\tree (D_s-2)&\frac{(p^2)^2}{2(p^2 - m^2)^4}\frac{(D-2)(D-4)}{D(D+2)} \\
&\ \ \ \times \pol_1^{\mu_1}\pol_2^{\mu_2}\pol_3^{\mu_3}\pol_4^{\mu_4}
(\eta_{\mu_1\mu_2}\eta_{\mu_3\mu_4}
+\eta_{\mu_1\mu_3}\eta_{\mu_2\mu_4}
+\eta_{\mu_1\mu_4}\eta_{\mu_2\mu_3})
\,,
\end{align}
which vanishes in $D=4$.  While in this case, the cancellation is
understood to be a consequence of supersymmetry~\cite{OneLoopSugraDiv},
it does provide a robust example illustrating that enhanced
cancellations become visible in the amplitudes only after making use of integral
identities.


\subsection{Two-loop example}

Enhanced cancellations become more interesting beyond one loop where
they correspond to a variety of ultraviolet cancellations for which
standard-symmetry explanations are not
known~\cite{VanishingVolume,KellyAttempt,HalfMaxMatter}.  We therefore
turn to half-maximal supergravity at two loops.  In $D=4$ the
cancellations are well understood to be a consequence of
supersymmetry~\cite{TwoLoopSugraDiv}, but in $D=5$ no such explanation
is known~\cite{TwoLoopHalfMaxD5}.

In $D=4$ we can enormously simplify the integrand by using helicity
states.  A simple trick that helps us simplify the analysis in higher
dimensions as well is to start with the higher-dimensional theory but
to restrict the external states and momenta to live in a
four-dimensional subspace.  In this way we can use four-dimensional
helicity methods to enormously simplify higher-dimensional integrands
as well.  This trick, of course, does not work for all states in the
higher-dimensional theory, but is sufficient for our purpose of
illustrating the difficulty of exposing enhanced cancellations at the
integrand level.

\begin{figure}[tb]
\captionsetup[subfigure]{labelformat=empty}
\begin{center}
\subfloat[\large (a)]{{\includegraphics[scale=.42]{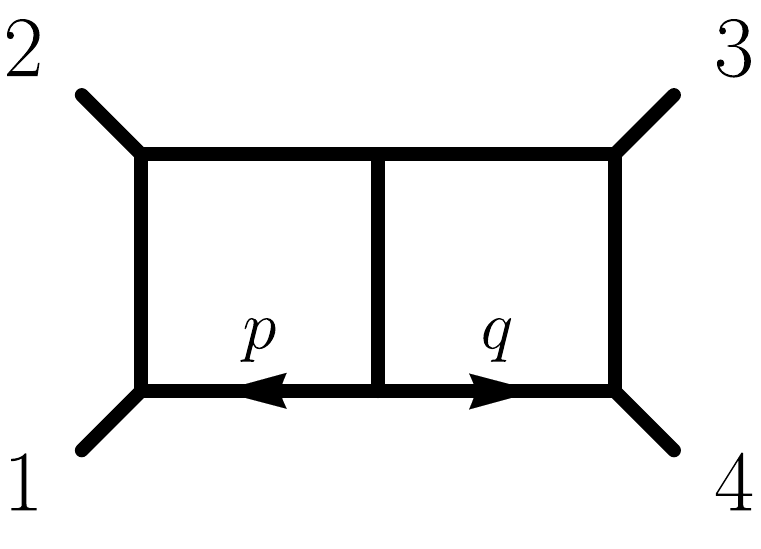}}}
  \label{subfig:planarDB}
\hspace{1cm}
\subfloat[\large (b)]{\includegraphics[scale=.42]{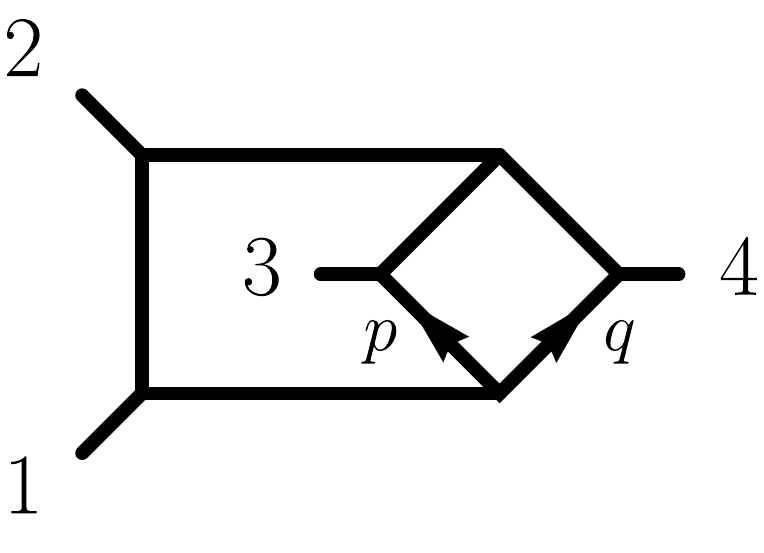}}
  \label{subfig:nonplanarDB}
\end{center}
\vskip -.3cm
\caption[a]{\small The planar and nonplanar double-box diagrams
that contribute to the four-point amplitudes of $\NeqFour$ supergravity.}
\label{fig:diags}
\end{figure}

Consider the four-point two-loop amplitude of $\NeqFour$ supergravity.
This amplitude has already been discussed in some detail in
Ref.~\cite{DixonTwoLoopN4}.  The double-copy construction of the
two-loop integrand is rather straightforward. We start from the
dimensionally-regularized $D=4$
all-plus helicity $(++++)$ pure Yang--Mills amplitude in the form
given in Ref.~\cite{ColorKinOneTwoLoops}.  (An earlier form of the
integrand may be found in Ref.~\cite{AllPlusTwoLoop}.) In this
representation the kinematic numerators of the planar and nonplanar double-box diagrams shown in \fig{fig:diags} are
\begin{align}
&n^{\mathrm{P\, YM}}_{1234}= {\cal T} \Bigl((D_s-2)s\left(\lambda_p^2\lambda_q^2+
\lambda_p^2\lambda_{p+q}^2+\lambda_q^2\lambda_{p+q}^2\right) + 16s\left((\lambda_p\cdot\lambda_q)^2
  -\lambda_p^2\lambda_q^2\right) \nonumber\\
&\hspace{1.5cm}+\frac{1}{2}(D_s-2)(p+q)^2\left((D_s-2)
  \lambda_p^2\lambda_q^2+8\left(\lambda_p^2+\lambda_q^2\right)
  \left(\lambda_p\cdot\lambda_q\right)\right) \Bigr)\,, \label{eq:allPlusPlanar} \\
&n^{\mathrm{NP\, YM}}_{1234} = {\cal T}\Bigl((D_s-2)s
 \left(\lambda_p^2\lambda_q^2+\lambda_p^2\lambda_{p+q}^2
 +\lambda_q^2\lambda_{p+q}^2\right)
 +16 s \left((\lambda_p\cdot\lambda_q)^2
 -\lambda_p^2\lambda_q^2\right) \Bigr)\,,
\label{eq:allPlusNonPlanar}
\end{align}
where $D_s$ is the state-counting parameter similar to that at one
loop and the subscript `1234' refers to the diagram external leg
labeling as in \fig{fig:diags}.  The momenta $p$ and $q$ are the
momenta carried by the propagators indicated in \fig{fig:diags}, while
$\lambda_p$ and $\lambda_q$ are their $(-2 \eps)$ components, where
$\eps = (4-D)/2$.  We use $\lambda_{p+q}$ as a shorthand for
$\lambda_p + \lambda_q$.
The crossing symmetric prefactor
\begin{equation}
{\cal T} = \frac{[12][34]}{\langle 12 \rangle \langle 34 \rangle}\,,
\end{equation}
is defined in terms of spinor inner products, following the notation
of Ref.~\cite{ManganoParke}. The remaining planar and nonplanar
double-box numerators are given by relabeling these. There are
contributions to the Yang--Mills integrand from other types of diagrams
as well, but we will not need them for the double-copy procedure.

To obtain half-maximal supergravity we then take the pure-Yang--Mills
amplitude and replace the color factors with $\NeqFour$
super-Yang--Mills numerators that satisfy BCJ duality using
\eqn{BCJDuality}. For the two-loop four-point amplitude of
$\NeqFour$ SYM a representation that satisfies the duality happens to
match the original construction~\cite{BRY}. The only nonvanishing
diagrams are the planar and nonplanar double boxes shown in
\fig{fig:diags}. The substitution (\ref{ColorSubstitute}) is simply
\begin{align}
 c^{{\rm P\,}}_{1234} &\rightarrow n^{{\rm P\,} \NeqFour}_{1234}
  = s^2 t A^\tree_{\NeqFour}(1,2,3,4)\,,
\nonumber \\
c^{\rm NP}_{1234} &\rightarrow n^{{\rm NP\,} \NeqFour}_{1234}
  = s^2 t A^\tree_{\NeqFour}(1,2,3,4)\,,
\end{align}
where numerators other than the planar and nonplanar ones vanish.  As
for the one-loop case, we package the $\NeqFour$ super-Yang--Mills
tree amplitude for all states into a single superamplitude.  The
half-maximal supergravity amplitude is then obtained by summing over
the planar and nonplanar double boxes in \fig{fig:diags}, with
kinematics numerators given by the product of pure Yang--Mills and
$\NeqFour$ super-Yang--Mills numerators,
\begin{align}
N^{\rm P \, half\hbox{-}max.\, sugra}_{\rm 1234}
& = s^2 t A^\tree_{\NeqFour}(1,2,3,4) \times
n^{\mathrm{P\, YM}}_{1234}\,, \nonumber \\
N^{\rm NP \, half\hbox{-}max. \, sugra}_{\rm 1234}
& = s^2 t A^\tree_{\NeqFour}(1,2,3,4) \times
n^{\mathrm{NP\, YM}}_{1234}\,.
\label{TwoLoopDoubleCopy}
\end{align}
The remaining supergravity planar and nonplanar double-box
numerators are given by simple relabelings. Diagrams other than the
planar and non-planar double boxes vanish.

This construction is also valid for the $D=5$ theory with the external
states restricted to a $D=4$ subspace.  We simply take $\eps
\rightarrow -1/2 +\eps$ and accordingly the $\lambda_p$ and
$\lambda_q$ become one dimensional up to $\mathcal{O}(\eps)$
corrections.  Similarly the state-counting parameter should be
shifted, $D_s \rightarrow D_s + 1$.  With these modifications, the
simple integrand in \eqn{TwoLoopDoubleCopy} is valid for the $D=5$
theory as well.

As terminology for the rest of the paper, when we label an amplitude
by its external helicity, we are not referring to the helicities of
the supergravity theory, but to the helicities of the pure Yang--Mills
theory comprising one side of the double-copy supergravity theory.

\subsubsection{Cuts and labels for nonplanar amplitudes}

Enhanced cancellations generally occur between diagrams of different
topologies.  A difficulty for exposing the cancellations at the
integrand level beyond one loop is that there is no unique and
well-defined notion of an integrand involving nonplanar diagrams.  Nor
is it clear in general how one should choose momentum labels in each
diagram that would allow cancellations between diagrams of various
topologies to occur.  For planar diagrams there is a canonical choice
of global variables for all diagrams based on dual
variables~\cite{NimaAllLoops}, but no analogous notion is known in the
nonplanar case.  As a simple example consider the planar and nonplanar
double-box diagrams in \fig{fig:diags}.  Fundamentally, the propagator
structure is different, making it unclear how one might be able show
the cancellation without integration.

A way to sidestep the labeling issue is to focus on unitarity cuts.
Generalized unitarity cuts that place at least one line on-shell in
every loop impose global momentum labels on the cut.  We can then ask
whether we can find nontrivial cancellations in the cut linked to enhanced
cancellations.  If such cancellations happen at the level of the
integrand, one should expect an improvement in the overall power
counting after summing over all contributions to the cuts compared to
individual terms.  Some care is required because cuts can also obscure
cancellations by restricting the diagrams that appear.  The
more legs that are cut, the fewer diagrams are included, since only
those diagrams that contain propagators corresponding to the cut ones
will be included.  Because of this, it is best to focus on cuts where
only a few legs are placed on shell.

\subsubsection{Absence of cancellations in a three-particle cut}

\begin{figure}[tb]
  \centering
  \includegraphics[width=0.42\textwidth]{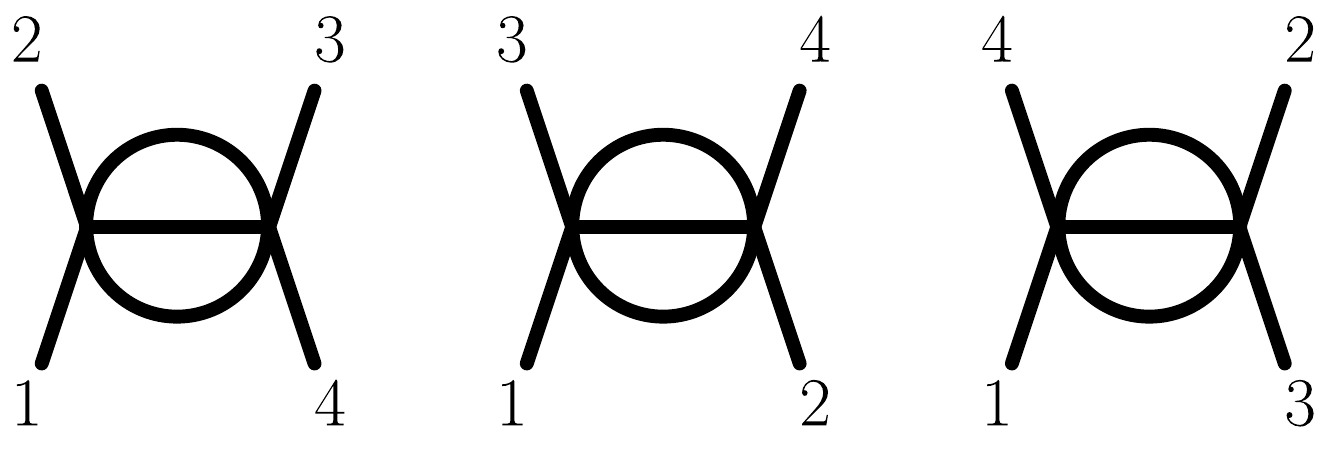}
  \caption{The three sunset integrals. These are ultraviolet divergent in $D=4$ and
    $D=5$.}
  \label{fig:sunsets}
\end{figure}

\begin{figure}[tb]
\begin{align}
\includegraphics[scale=0.42,trim= 0 80 5 0]{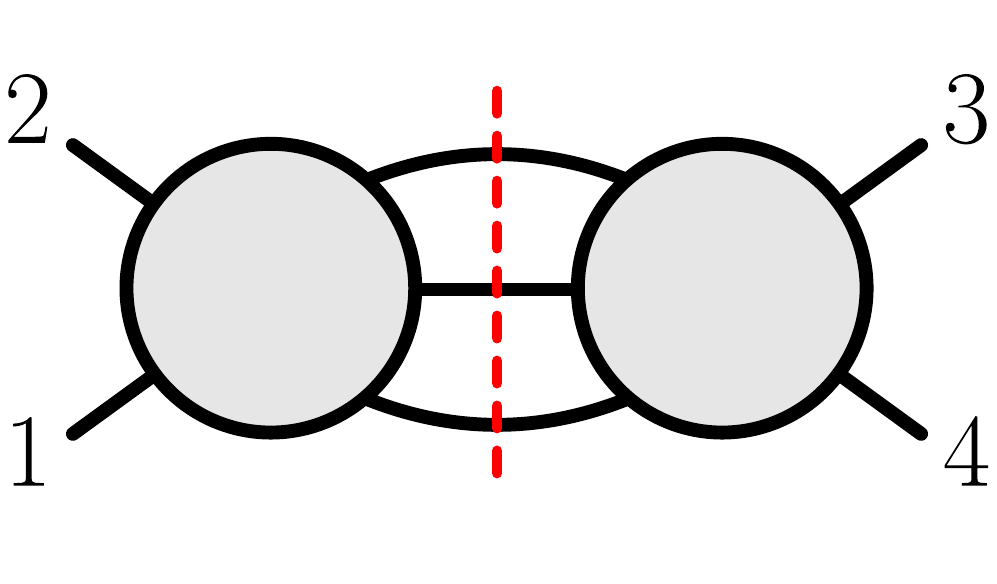}   \nonumber
  &= \frac{1}{2} \times  \includegraphics[scale=0.4,trim= 20 117 0 0]{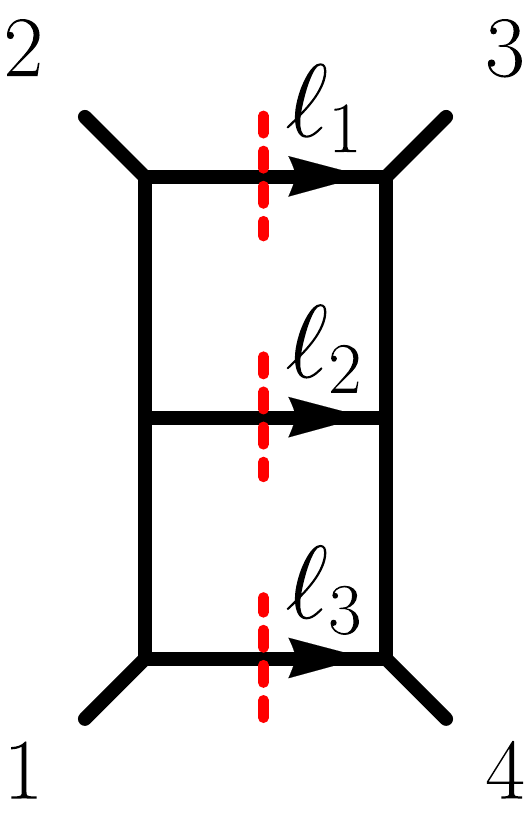}
  + \includegraphics[scale=0.4,trim=0 80 0 0]{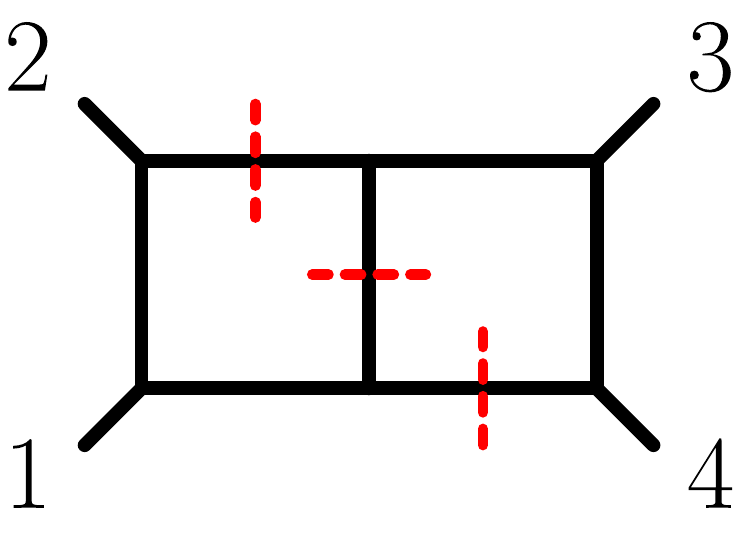}   \\ \nonumber
  &\\ \nonumber
  &\\ \nonumber
  & \hspace{-4cm} + \includegraphics[scale=0.4,trim=20 80 0 0]{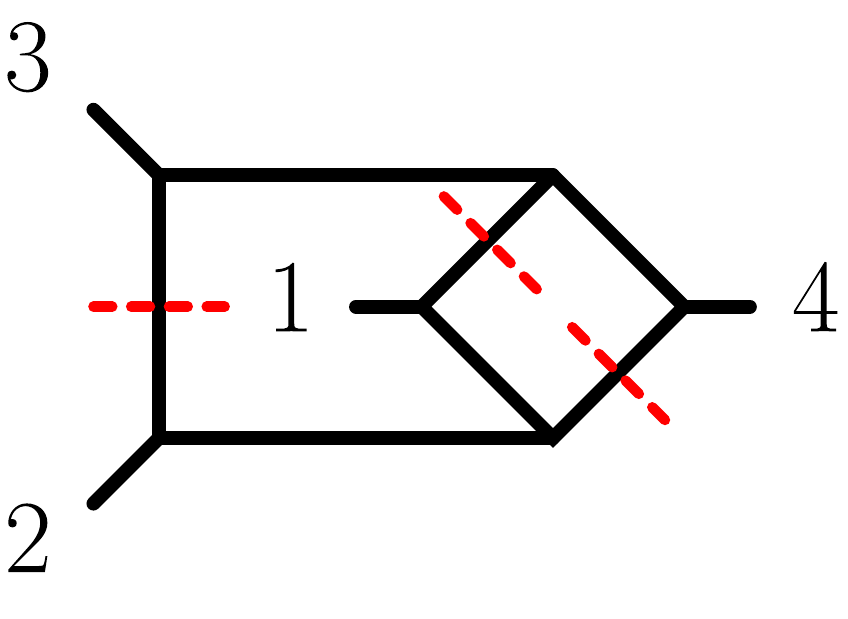}
  + \frac{1}{2} \times \includegraphics[scale=0.4,trim=20 80 0 0]{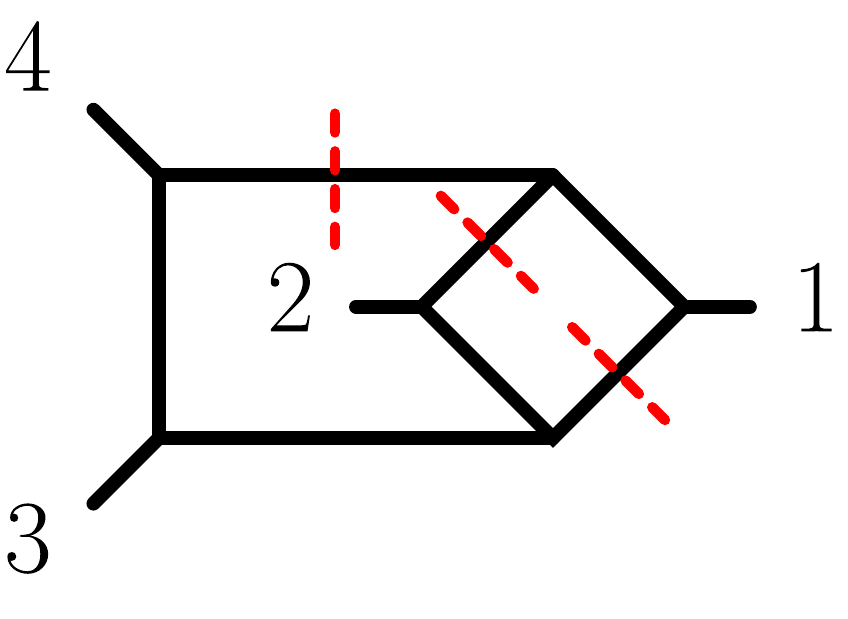}
  + \frac{1}{2} \times \includegraphics[scale=0.4,trim=20 80 0 0]{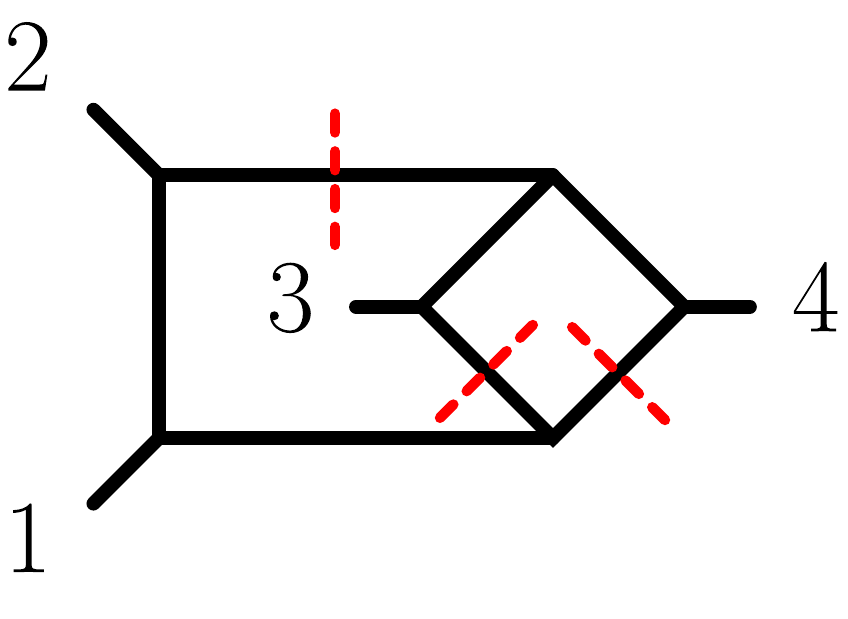}  \\ \nonumber
  & \\ \nonumber
  & \\
  & \hspace{-2cm} + \; \text{perms}(\ell_1, \ell_2, \ell_3) \;
      + \;  (1 \leftrightarrow 2)\;
      +  \; (3 \leftrightarrow 4) \;
      + \;  (1 \leftrightarrow 2\;,\; 3 \leftrightarrow 4). \nonumber
\end{align}
\caption{The contributing diagrams of the three-particle cut of
        the half-maximal supergravity two-loop four-point amplitude.
        The shaded (red) dashed lines indicated the legs which are cut.  }
\label{fig:3pcut}
\end{figure}

The three-particle cut in \fig{fig:3pcut} is useful for studying
enhanced cancellations.  In the following section, using
integration-by-parts technology we describe an arrangement
of the integrand where potential divergences are pushed into
sunset diagrams, illustrated in \fig{fig:sunsets}.
This suggests that the three-particle cut, where the cut
lines correspond to the three propagators of a sunset diagram, is a
natural one for studying enhanced cancellations. In addition, this cut
fixes all loop momentum labels in this amplitude in terms of the
momenta of the cut lines.  An obvious guess is that if we apply the
three-particle cut corresponding to the internal lines of the sunset
diagram, we should find improved power counting in the full sum over
terms compared to individual contributions.

The $(++++)$ amplitude has a number of special features that
simplify the analysis of the cut, making it easier to find
ultraviolet cancellations if they exist.  On the three-particle cut,
the terms in the numerator proportional to $(p+q)^2$ in
\eqn{eq:allPlusPlanar} are set to zero because they corresponds to one
of the on-shell inverse propagators $\ell_1^2$, $\ell_2^2$ or
$\ell_3^2$, as can be seen in \fig{fig:3pcut}, making the form of the
planar and nonplanar numerators identical in the three-particle cut. A
useful feature of the remaining numerator terms that we exploit is
that they are invariant under relabelings: the expression is the same
under any mapping of the $p$ and $q$ propagator labels to any two of
the three $\ell_1$, $\ell_2$ and $\ell_3$.  In addition, up to
prefactors depending on external momenta, the dependence of the
numerators is only on the components outside the four-dimensional
subspace where the external momenta and helicities live.  These
features enormously simplify the analysis of the cut because most of
the numerator factors out and is independent of permutations of
external or internal legs.

Using these observations, after inserting the numerators into the
planar and nonplanar double-box diagrams and taking the three-particle
cut shown in \fig{fig:3pcut}, we obtain the expression:
\begin{align}
 \nonumber \mathcal{I}^{\text{cut}}&= {\cal P}(\ell_1,\ell_2,\ell_3)
 \\ \nonumber
\times & \biggl[ \Bigl(  \nonumber
 \frac{1}{2}\, \frac{t^2}{(\ell_1+k_1)^2 (\ell_3+k_2)^2 (\ell_3-k_3)^2
   (\ell_1-k_4)^2} + \frac{t^2}{(\ell_2+k_1)^2 (\ell_3+k_2)^2
  (\ell_3-k_3)^2 (\ell_1-k_4)^2} \\ \nonumber
 & +\frac{s^2}{(\ell_1+\ell_2)^2 (\ell_2+\ell_3)^2
  (\ell_3+k_2)^2 (\ell_1-k_4)^2}
 +\frac{1}{2} \,
 \frac{s^2}{(\ell_2+\ell_3)^2 (\ell_3+k_1)^2 (\ell_2+k_2)^2
          (\ell_1-k_4)^2}   \\ \nonumber
 &  +{1\over 2}\,
 \frac{s^2}{(\ell_1+\ell_2)^2 (\ell_3+k_2)^2 (\ell_2-k_3)^2 (\ell_1-k_4)^2}
 \Bigr) + \hbox{perms}(\ell_1, \ell_2, \ell_3) \\
 & \hskip 5 cm
  + (1 \leftrightarrow 2 )
    + ( 3 \leftrightarrow 4)
  + (1 \leftrightarrow 2 ,\, 3 \leftrightarrow 4)  \biggr]\,,
\label{ThreePartCut}
\end{align}
where the on-shell conditions $\ell_1^2 = \ell_2^2 = \ell_3^2 = 0$ are imposed.
The prefactor ${\cal P}(\ell_1,\ell_2,\ell_3)$ is
\begin{align}
{\cal P}(\ell_1,\ell_2,\ell_3)= & -i (D_s-2) s t A^\tree_{\NeqFour}(1,2,3,4)\, {\cal T}
 \nonumber \\
& \null \times
\Bigl( \left(\lambda_{\ell_1}^2\lambda_{\ell_2}^2
 +\lambda_{\ell_1}^2\lambda_{\ell_3}^2
 + \lambda_{\ell_2}^2\lambda_{\ell_3}^2\right)
 + 16 s \left((\lambda_{\ell_1}\cdot\lambda_{\ell_2})^2
 - \lambda_{\ell_1}^2\lambda_{\ell_2}^2\right) \Bigr)\,,
 \label{PreFactor}
\end{align}
which is invariant under the permutations of external and internal cut
legs indicated in \eqn{ThreePartCut}.  We have
analyzed \eqn{ThreePartCut} both analytically and numerically and we
find that for $\ell_i \rightarrow \infty$ there is no improvement in
the large loop-momentum behavior after summing over all terms,
compared to the behavior of a single term.  In fact, this is no
surprise because other than the overall prefactor (\ref{PreFactor}),
this sum over terms is {\it precisely} the same one that appears in
the three-particle cut of the two-loop four-point amplitude of
$\NeqEight$ supergravity given in Eq.~(5.15) of Ref.~\cite{TwoLoopN8}.
In $\NeqEight$ supergravity we know there are no further cancellations
arising from the sum over diagrams.  This can be seen as follows: the
only nonvanishing diagrams in $\NeqEight$ supergravity are the planar
and nonplanar double boxes of \fig{fig:diags}, but with no loop
momenta in the numerators~\cite{TwoLoopN8}.  Simple power counting
shows that each diagram of $\NeqEight$ supergravity is ultraviolet
divergent in dimensions $D\ge 7$.  This divergence does not cancel in
the sum over diagrams, leading to a divergence of the four-point
amplitude of $\NeqEight$ supergravity~\cite{TwoLoopN8}:
\begin{equation}
{\cal M}_4^{\twoloop, D=7-2\eps} \Bigl|_{\rm UV\ div.} =
\frac{1}{2 \eps (4 \pi)^7} \frac{\pi}{3} (s^2 + t^2 + u^2) \times
\biggl(\frac{\kappa}{2}\biggr)^6  \, stu M^\tree_{{\NeqEight}}(1,2,3,4)\,,
\end{equation}
where we have stripped the coupling constant and
$M^\tree_{{\NeqEight}}$ is the supergravity tree amplitude.  The fact
that there are no further cancellations in $\NeqEight$ supergravity
implies that no integrand-level cancellation is possible in our
$\NeqFour$ supergravity three-particle cut (\ref{ThreePartCut}).  One
might imagine trying to include relabelings $\ell_i \rightarrow
-\ell_i$ in the spirit of Ref.~\cite{EnricoJaraIR} or other
relabelings in order to try to expose cancellations.  However, because
of the link to the $\NeqEight$ supergravity cut, it is clear there are
no further cancellations to be found.

In summary, we see no evidence of cancellations at the integrand
level.  The usual supergraph Feynman rules or amplitudes-based proofs
of ultraviolet finiteness in gauge theory (see for example,
Ref.~\cite{Finiteness}) rely on the ability to make the integrand
manifestly ultraviolet finite by power counting.  The difficulty in
finding a standard-symmetry based explanation for enhanced
cancellations~\cite{VanishingVolume,KellyAttempt,HalfMaxMatter} in
gravity theories is presumably tied to our difficulty in identifying
the cancellations at the integrand level.  This greatly complicates
any all-order understanding of the divergence properties of
supergravity theories. If we are to unravel enhanced cancellations, we
need to turn to the systematics of cancellations from integral
identities.


\section{Rearranging the integrand to show finiteness}
\label{sec:rearranging}

As discussed in the previous section, it does not appear possible to
expose enhanced cancellations purely at the integrand level. In this
section we show how one can rearrange integrands into a form where all
terms are manifestly finite by power counting, except those that
integrate to zero.  We do so using modern integration-by-parts (IBP)
technology~\cite{IBP, KosowerIBP, ItaIBP, LarsenZhang}.  In our discussion we will
be using the language of integrands and integrals
interchangeably. This is because the modern approaches to integration
by parts can be used to track terms in the integrand that integrate to
zero, in a manner analogous to the one-loop technology of
Refs.~\cite{OPP,Forde}.

We first outline how IBP relations can be used to reorganize
integrands with enhanced cancellations so that all terms that are
naively ultraviolet divergent by power counting integrate to zero.  We
start from a given integrand that has the schematic structure
\begin{equation}
  \mathcal{I}^{\text{total}}=\sum_i \mathcal{I}_i^{\text{fin.}}
  + \sum_j \mathcal{I}_j^{\text{div.}}\,.
  \label{genericRep}
\end{equation}
The sum runs over the various pieces of the integrand, denoted
by $\mathcal{I}_i^{\text{fin.}}$, which are finite by power counting,
and $\mathcal{I}_i^{\text{div.}}$ which are divergent by power
counting. After integration, however, the total may be
finite. The idea is to reorganize this integrand into the form
\begin{equation}
  \mathcal{I}^{\text{total}}=\sum_i \tilde{\mathcal{I}}_i^{\text{fin.}}
                + \sum_j \tilde{\mathcal{I}}_j^{\text{van.}}\,,
  \label{finiteRep}
\end{equation}
where $\tilde{\mathcal{I}}_i^{\text{fin.}}$ is another set of integrands that
are finite after integration and $\tilde{\mathcal{I}}_j^{\text{van.}}$ can be
divergent by power counting but integrate to zero,
\begin{equation}
  \int \tilde{\mathcal{I}}_j^{\text{van.}} = 0\,,
\end{equation}
thus making the finiteness manifest.  The reorganization is accomplished by
writing the sum over power-counting divergent integrals as
\begin{equation}
\sum_j \mathcal{I}_j^{\text{div.}} = \sum_j\mathcal{I'}_j^{\text{fin.}} +
         \sum_j (\mathcal{I}_j^{\text{div.}} - \mathcal{I'}_j^{\text{fin.}}) \,,
\end{equation}
where the terms in parentheses integrates to zero and the finite
integrals $\mathcal{I'}_j^{\text{fin.}}$ are included with the
finite ones in \eqn{finiteRep}.

IBP technology offers a systematic means for accomplishing this.  We
briefly review this.  The IBP method~\cite{IBP} takes
advantage of the fact that in dimensional regularization a total
derivative vanishes:
\begin{equation}
\int \prod_i d^D\ell_i \, \frac{\partial}{\partial \ell_j^{\mu}}
 \left(\frac{v_j^\mu}{\prod_k D_k}\right) =0\,,
 \label{IBPe}
\end{equation}
where $1/D_k$ are propagators and $v^{\mu}_j$ are arbitrary functions
of loop momenta as well as external kinematics or other vectors in the
problem.  Evaluating
the derivatives gives a sum of terms, and the vanishing of the integral
therefore implies a relation amongst the integrals corresponding to
each term.  By exhausting all such independent relations one can
choose a basis of integrals in terms of which to express a given
amplitude. The standard basis choice at one loop is a combination of
boxes, triangles, and bubbles~\cite{PassarinoVeltman}, but at higher
loops there is no canonical choice.  In general, different
bases might be used to manifest different aspects of the amplitude,
such as its symmetries and/or behavior on certain unitarity cuts.

Generically, when applying integration-by-parts identities, there is
no natural separation of the type in \eqn{finiteRep}.  In general, the
coefficients of individual terms can develop $1/\epsilon$
singularities, and divergences cancel in complicated ways, making the
finiteness unclear.  To avoid this, some care is required to pick
integral bases that (a) do not introduce divergences in integral
coefficients and (b) contain a minimal number of divergent integrals.
Usually, one picks a linearly independent set of integrals, because
this minimizes the number of objects that need to be computed.  But,
even for an ultraviolet finite amplitude, a general choice of basis
will likely have explicit ultraviolet divergences either in basis
integrals or in their coefficients.  The finiteness is thus obscured
because the divergence cancels only in the full sum over
contributions.  A way to avoid this problem and express the amplitude
in the form of \eqn{finiteRep} is to use an overcomplete set of
integrals.  The overcompleteness gives sufficient freedom that we can
exploit to make the finiteness manifest.

We illustrate this procedure with a simple example.
Suppose our expression is given as the sum of integrals:
\begin{align}
A &=\frac{1}{70} \includegraphics[scale=0.4,trim= 0 55 0 0]{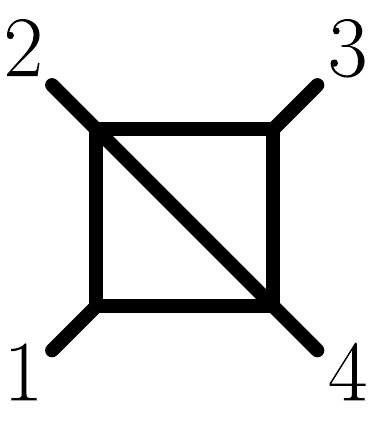}
      - \frac{1}{2s^2} \includegraphics[scale=0.4,trim= 0 55 0 0]{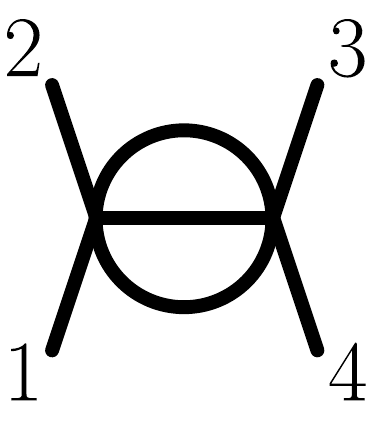}
-\frac{1}{2t^2} \includegraphics[scale=0.4,trim= 0 55 0 0]{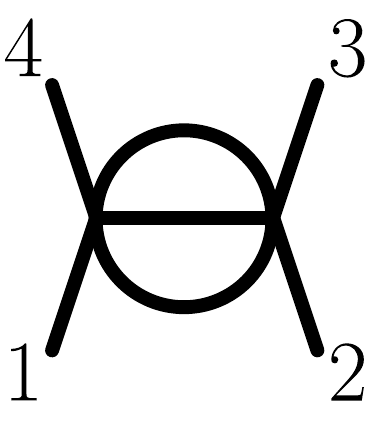}.
 \label{toyAmp} \\ \nonumber
\end{align}
Each of these integrals are ultraviolet divergent in five dimensions
with the following leading divergences (omitting an overall $\pi/32$):
\begin{align}
\includegraphics[scale=0.4,trim= 0 55 0 0]{figures/dbox.pdf}\Bigg|_{\rm UV\ div.}&=\frac{1}{3 \epsilon}\,, \qquad
     \includegraphics[scale=0.4,trim= 0 55 0 0]{figures/sset1234.pdf}\Bigg|_{\rm UV\ div.}=\frac{s^2}{210 \epsilon}\,,  \qquad
     \includegraphics[scale=0.4,trim= 0 55 0 0]{figures/sset1432.pdf}\Bigg|_{\rm UV\ div.}=\frac{t^2}{210 \epsilon}\,.
 \\ \nonumber
\end{align}
Evaluating the divergence shows that \eqn{toyAmp} is finite, but this is not
manifest in the above representation.  Now consider the following IBP
identities
\begin{align}
  d \omega_1 &= \includegraphics[scale=0.4,trim= 0 55 0 0]{figures/dbox.pdf}
      - \frac{70}{s^2} \includegraphics[scale=0.4,trim= 0 55 0 0]{figures/sset1234.pdf}
 - \frac{1}{3 s^2} \includegraphics[scale=0.4,trim=0 55 0 0]{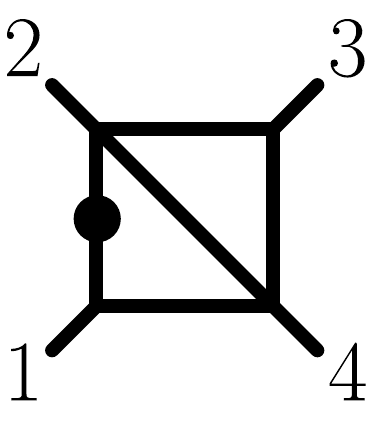}\,, \nonumber \\
  \nonumber & \\
  d \omega_2 &= \includegraphics[scale=0.4,trim= 0 55 0 0]{figures/dbox.pdf}
      + \frac{70}{su}\includegraphics[scale=0.4,trim= 0 55 0 0]{figures/sset1234.pdf}
 + \frac{70}{tu} \includegraphics[scale=0.4,trim= 0 55 0 0]{figures/sset1432.pdf}
 - \frac{s t}{3 u} \includegraphics[scale=0.4,trim=0 55 0 0]{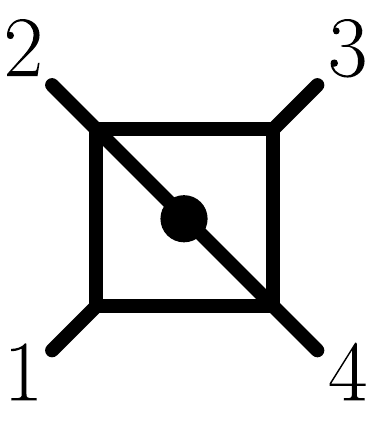}\,, \label{IBP}
 \\   \nonumber &
\end{align}
where $d\omega_1$ and $d\omega_2$ are appropriate total derivatives; their
precise form is not important for our purposes.  The dot placed on a
propagator indicates that the propagator is doubled, i.e., squared.
This choice is convenient because the two integrals with doubled
propagators are both ultraviolet finite in $D=5$.

For this simple example, one can solve this system of equations for
two of the three ultraviolet-divergent integrals.  Plugging in the
solution leaves only a single ultraviolet-divergent integral whose coefficient
must vanish, if the amplitude is finite.  However, the ability to
express $A$ in \eqn{toyAmp} in terms of a basis of manifestly finite integrals is a
consequence of the simplicity of this example, and for more
complicated amplitudes this straightforward approach will not suffice.
We will therefore take a more general approach for this example.
In particular, we can use \eqn{IBP} to rewrite the crossed box
integral as
\begin{align}
  \includegraphics[scale=0.4,trim= 0 55 0 0]{figures/dbox.pdf}&=
  \alpha \Big( - \frac{70}{su}\includegraphics[scale=0.4,trim= 0 55 0 0]{figures/sset1234.pdf}
- \frac{70}{tu} \includegraphics[scale=0.4,trim= 0 55 0 0]{figures/sset1432.pdf}
+ \frac{s t}{3 u} \includegraphics[scale=0.4,trim=0 55 0 0]{figures/dboxc.pdf} \Big) \nonumber \\
  \nonumber & \\
  & \ \ \  + (1-\alpha) \Big(  \frac{70}{s^2} \includegraphics[scale=0.4,trim= 0 55 0 0]{figures/sset1234.pdf}  + \frac{1}{3 s^2} \includegraphics[scale=0.4,trim=0 55 0 0]{figures/dboxl.pdf}  \Big) +  d\Bigl( (1-\alpha)\,\omega_1 + \alpha\,\omega_2 \Bigr) \,,
 \label{overComplete}
 \\   \nonumber &
\end{align}
where $\alpha$ is a free parameter.  In this way we traded one
ultraviolet-divergent integral for two ultraviolet-divergent sunset
integrals which were already in the basis, plus two other finite
integrals and a collection of integrals that vanish (i.e., are total
derivatives).  Plugging this back into the original expression for $A$
gives
\begin{align}
A =  \Big( \frac{1-\alpha}{s^2}-\frac{\alpha}{su}-\frac{1}{2s^2} \Big) & \includegraphics[scale=0.4,trim= 0 55 0 0]{figures/sset1234.pdf}
-\Big( \frac{\alpha}{tu}+\frac{1}{2t^2} \Big)\includegraphics[scale=0.4,trim= 0 55 0 0]{figures/sset1432.pdf} \\ \nonumber    \\ \nonumber & +\text{finite}
  +\frac{1}{70}d\Bigl( (1-\alpha)\,\omega_1 + \alpha\,\omega_2 \Bigr)\,,
\end{align}
where ``finite'' corresponds to integrals that are manifestly ultraviolet
finite with finite coefficients and the term
$\frac{1}{70}d(...)$ vanishes upon integration.  For general
$\alpha$ this form of $A$ is still not manifestly finite, but since
$\alpha$ is arbitrary we can take it to be $\alpha=-{u}/{2t},$ in
which case the coefficients to the two sunsets both vanish and $A$ is
then manifestly a sum of finite integrals and integrals that vanish.
In general, one free parameter will not be enough to tune away two
coefficients of ultraviolet-divergent integrals.  For more complicated examples
one needs to generate more IBP relations and introduce more tunable
parameters, and in general each parameter can be used to set one
coefficient to an ultraviolet-divergent integral to zero.

As a nontrivial example, we carried out this procedure for the
$(-+++)$ two-loop amplitude of half-maximal supergravity in
$D=5$. (Recall that the helicity labels refer to the helicities of the
pure Yang--Mills side of the double copy, with the external states
restricted to live in a four-dimensional subspace.)  The structure of
this amplitude is much more complicated than the $(++++)$ case and
more representative of generic cases.  In the first step we reduce the
full integrand to a basis of master integrals using Larsen and Zhang's
method~\cite{LarsenZhang}.  After this procedure the only contributing
ultraviolet-divergent integrals are the three different labels of the
sunsets and a few others.  We then used these types of over-complete
relations to express all of the (non-sunset) ultraviolet-divergent
integrals in terms of ultraviolet-divergent sunset integrals, finite
integrals and total derivatives that integrate to zero.  The tunable
parameters are solved so that coefficients of the three sunsets vanish
separately, while maintaining finiteness of the coefficients of all
finite integrals.  Therefore, by allowing for an over-complete basis
and tuning the parameters that keep track of this over-completeness,
we are able to write the amplitude in the desired form,
Eq.~(\ref{finiteRep}).

We note that unless special care is taken, an IBP identity in general
involves doubled propagators, as in \eqn{overComplete}. This has the
unwanted side effect of introducing spurious infrared
singularities even in $D=5$. With more modern
approaches~\cite{KosowerIBP, ItaIBP, LarsenZhang} we can avoid the
appearance of such integrals. This is achieved by imposing
\begin{equation}
\sum_{j} v_j^\mu \frac{\partial}{\partial \ell_j^\mu} D_k = f_k \, D_k\,,
\end{equation}
on the $v_j^\mu$ and where $f_k$ has polynomial dependence on
Lorentz-invariant dot products of momenta.  We have also applied the
more modern approach and find similar results.

The procedure sketched above shows that the $D=5$ two-loop four-point
integrand of half-maximal supergravity can be rewritten in a form that
is manifestly finite, up to terms that integrate to zero.  However,
this procedure relies on the specific details of the integrand and
corresponding IBP relations.  It is also computationally difficult to
extend to higher loops.  Clearly, we need an approach where the
necessary identities can be derived from generic properties of loop
integrals.  We will describe such an approach in the next section.

\section{Vacuum expansion and systematics of ultraviolet cancellations}
\label{sec:ibp}

In this section we describe a systematic approach to understanding
enhanced cancellations, in a manner that appears to have an all-loop
generalization.  We continue to focus on the two-loop amplitudes of
half-maximal supergravity. The ultraviolet behavior is determined at
the integrand level by large values of loop momenta, or equivalently
small external momenta.  It is therefore natural to series expand the
integrand in this limit. Although this expansion has the unwanted
effect of losing contact with the unitarity cuts and introducing
spurious singularities, such as doubled propagators, it does have the
important advantage of focusing on the term directly relevant for the
ultraviolet behavior.  In general, we are interested in the
logarithmic divergences, so we series expand to the appropriate order
where the integrals become logarithmically
divergent in ultraviolet~\cite{Vladimirov,HalfMaxMatter}. (We note that while
dimensional regularization does not have direct access to power
divergences, such divergences become logarithmic simply by lowering
the dimension.)  This expansion generates a set of vacuum integrals.
For example, at two loops these integrals have the form
\begin{equation}
 \int {d^D p} \,{d^D q}\,
\frac{\mathcal N(p, q, {k_i})}{(p^2)^A (q^2)^B ((p+q)^2)^C}\,,
\end{equation}
where $A,B$ and $C$ denote the powers of the propagators.  In addition
to being ultraviolet divergent, these vacuum integrals also are
infrared divergent. This complicates the extraction of the ultraviolet
divergences. For example, in dimensional regularization these
integrals are scaleless, and the infrared singularities exactly cancel
the ultraviolet ones.  This is usually dealt with by introducing a
mass regulator or by injecting external momentum into the diagram.
(See, for example, Refs.~\cite{Vladimirov,Simplifying,HalfMaxMatter}.)
We will avoid this complication by systematically finding relations
between the divergences of the integrals using integration by parts.

As noted in the previous section, the simplest example to analyze is
the case where the external gluons in the pure Yang--Mills side of the
double-copy are restricted to live in four dimensions, and
correspond to all-plus helicity $(++++)$. For this helicity
configuration on the pure Yang--Mills side of the double copy, we use
the spinor-helicity integrands in
\eqns{eq:allPlusPlanar}{eq:allPlusNonPlanar}. For the remaining
helicity configurations we used the pure Yang--Mills integrand from
Ref.~\cite{DoubleCopyUnitarity}. The only contributions needed are
those whose color structure matches those of the planar and nonplanar
double-box diagrams. For other helicities we used the gauge-invariant
projection method to be described in Ref.~\cite{SuperGaussBonnet}.

In four-dimensions these integrals do not have overall ultraviolet
divergences because they are suppressed by the numerators; they are
proportional to the $(-2\eps)$-dimensional components of loop
momenta. (They do however contain subdivergences which cancel.)  To
have a nontrivial example, we turn to the same integrand but with the
internal states in $D=5$. In this case the numerator is not suppressed
because $\lambda_p$ and $\lambda_q$ are one-dimensional. (In the context of
dimensional regularization in $D=5-2\eps$, they are actually $(1-2\eps)$
dimensional.)  Using $D=5$ properties the integrand simplifies: In
$D=5$ the $\lambda_{p}$ and $\lambda_q$ become one-dimensional so that
\begin{equation}
(\lambda_p \cdot \lambda_q)^2 - \lambda_p^2 \lambda_q^2 \rightarrow \mathcal{O}(\eps)\,,
\end{equation}
in \eqns{eq:allPlusPlanar}{eq:allPlusNonPlanar}.

In the large loop-momentum limit, the logarithmically divergent terms
in $D=5$ are given by
\begin{equation}
  I^{\rm P,\, NP} =  (D_s-2) s  \int d^D p\, d^D q \,
 \frac{ \left( \lambda_p^2 \lambda_q^2 +
\lambda_p^2 \lambda_{p+q}^2 + \lambda_q^2 \lambda_{p+q}^2 \right)}
 {{(p^2)}^{A} {(q^2)}^{B} {[(p+q)^2]}^{C}}  + \text{UV finite} \,,
\label{eq:expanded}
\end{equation}
where
\begin{equation}
(A,B,C) = \left.
\begin{cases}
  (3,3,1)\,, & \text{P: planar double box\,,}\\
  (3,2,2)\,, & \text{NP: nonplanar double box\,.}
\end{cases}
\right.
\label{eq:PandNPvacuum}
\end{equation}
In the planar case there are power divergences coming from terms
proportional to $(p+q)^2$, which removes the middle propagator
generating a product of one-loop integrals.  Such terms do not give
rise to logarithmic divergences. (This is consistent with finiteness
of such integrals in dimensional regularization, which is sensitive
only to logarithmic divergences.)  We may then ignore such terms for
the purposes of trying to understand overall two-loop logarithmic
divergence.

One way to evaluate \eqn{eq:expanded} is to consider vacuum integrals
with numerators that are polynomial in $v_j\cdot p$ and $v_j \cdot
q$, where the $v_j$'s are a set of orthonormal basis vectors for the
five-dimensional momentum space. We have
\begin{equation}
v_5 \cdot p = \lambda_p\,, \hskip .7cm
v_5 \cdot q = \lambda_q\,, \hskip .7cm
\sum_j (v_j \cdot p) (v_j \cdot p) = p^2, \hskip .7 cm
\sum_j (v_j \cdot q) (v_j \cdot q) = q^2\,, \hskip .5 cm
\end{equation}
with appropriate factors of $i$ inserted for the metric signature.
Lorentz invariance then implies
\begin{equation}
  \text{UV finite } = \int d^D p \, d^D q \, v_i^{[\mu} v_j^{\nu]}
\left( p_\mu \frac{\partial}{\partial p^\nu} +
  q_\mu \frac{\partial}{\partial q^\nu} \right)
\frac{\mathcal N(v_k\cdot p, v_k \cdot q)}
{ {(p^2)}^{A} {(q^2)}^{B} {[(p+q)^2]}^{C} } \,,
\label{Lorentz}
\end{equation}
where the Lorentz indices $\mu$ and $\nu$ are antisymmetrized.
 By replacing $\mathcal N$
in the above equation by all possible monomials in $v_i \cdot p$ and
$v_i \cdot q$ up to degree four,
we generate linear relations between vacuum integrals with different
numerators, allowing us to reduce \eqn{eq:expanded} to scalar vacuum
integrals.  The result of this procedure is
\begin{align}
  I^{\rm P,\,NP} &= \frac{3}{70} (D_s-2) s \int d^D p \, d^D q \,
 \frac{\left[ (p^2)^2 + (q^2)^2 +((p+q)^2)^2 \right]} { {(p^2)}^{A} {(q^2)}^{B} {[(p+q)^2]}^{C} }
    \nonumber \\
  &= \frac{3}{70}\,  (D_s-2) s (I_{A-2,B,C} + I_{A,B-2,C} + I_{A,B,C-2}) \,,
\label{eq:allPlusReduced}
\end{align}
where the scalar vacuum integrals are defined as
\begin{equation}
I_{A,B,C} = \int d^D p\, d^D q \frac{1}
  { {(p^2)}^{A} {(q^2)}^{B} {[(p+q)^2]}^{C} }\,,
\label{eq:twoLoopVac}
\end{equation}
which is invariant under the six permutations of $\{A,B,C\}$.
 One can also
obtain this equation by reducing the implicit tensor integrals
in \eqn{eq:expanded}, using Lorentz invariance in the more
traditional way following for example Eq.~(4.18) of Ref.~\cite{Simplifying}.
Alternatively, Mastrolia et.\ al.\ recently proposed an efficient
algorithm to integrate away loop momentum components orthogonal to all
external momenta~\cite{Mastrolia}.

For the particular cases of \eqn{eq:allPlusReduced} we obtain
\begin{align}
  I^{\rm P} &= \frac{3s}{70} (D_s -2 )
\left(   I_{1,3,1} + I_{3,1,1} + I_{3,3,-1}\right) +
 \text{UV finite} \nonumber \\
  &= \frac{3s}{70} (D_s -2 ) (2 I_{3,1,1} + I_{3,3,-1}) + \text{UV finite},
  \label{eq:vaccumnum1} \\
I^{\rm NP} &= \frac{3s}{70} (D_s -2 ) (I_{1,2,2}+I_{3,0,2}+I_{3,2,0})+ \text{UV finite} \, ,
\end{align}
where we used the fact that the integrals are invariant under the
exchange of $p$ and $q$ in the second equality in \eqn{eq:vaccumnum1}.
Summing the planar and nonplanar contributions, we conclude that the logarithmic UV divergence is given by
\begin{equation}
(I^{\rm P} +  I^{\rm NP})\big |_{\rm log\ UV} =
\frac{3s}{70} (D_s -2 ) \left( 2 I_{3,1,1} + I_{1,2,2} \right)\big |_{\rm log\ UV} \, .
\label{eq:PandNPdiv}
\end{equation}
As explained above, the terms with ``one-loop squared'' propagator structures (e.g., $I_{3,2,0}$ or $I_{3,3,-1}$ ) do not contain logarithmic UV divergences. Also, it is not surprising that the final result is a linear combination
of $I_{3,1,1}$ and $I_{1,2,2}$, as these are the only two possible
logarithmically divergent vacuum integrals in $D=5$.

By explicit evaluation using a uniform internal mass $m$ as an
infrared regulator and dimensional regularization in $5-2\epsilon$
dimensions as an ultraviolet regulator, we find
\begin{align}
I_{3,1,1}\Bigl|_{\rm UV\ div.} &= -\frac{\pi}{192 \epsilon} \,,
\nonumber \\
I_{1,2,2}\Bigl|_{\rm UV\ div.} &= \frac{\pi}{96 \epsilon}\,,
\label{eq:divs}
\end{align}
so the combination of integrals in Eq.\ \eqref{eq:PandNPdiv} is
ultraviolet finite in $D=5$.  However, in order to understand the general
structure of the cancellations, it is illuminating to instead show this
using IBP identities.

\subsection{Extracting divergences using IBP identities}
We recall that the fundamental assumption of the IBP method is that
the integral of a total derivative vanishes in dimensional
regularization, as shown in \eqn{IBPe}. Obviously, integrals of total
derivatives only vanish when boundary contributions vanish. In
dimensional regularization however, we can consider the integral in a
dimension where the boundary contribution is vanishing and then
analytically continue the result (zero) to the original dimension. But
in an another regularization scheme one has to consider the
behavior of boundary terms. In particular, if the boundary term
contains ultraviolet or infrared divergences itself, the corresponding IBP identity
cannot be used to relate the divergences of the integrals.

On the other hand, dimensional regularization is known to regulate the
ultraviolet and infrared simultaneously. In general this is very convenient, but this
fact might obstruct the use of certain IBPs in this scheme for
extracting ultraviolet divergences. The reason for this is that IBP identities
in dimensional regularization can mix up ultraviolet and infrared poles. To
illustrate this consider the following identity that relates
bubble and triangle integrals in $D=4$:
\begin{align}
d \omega = s\,\epsilon\times
\includegraphics[scale=0.5,trim= 20 60 5 0]{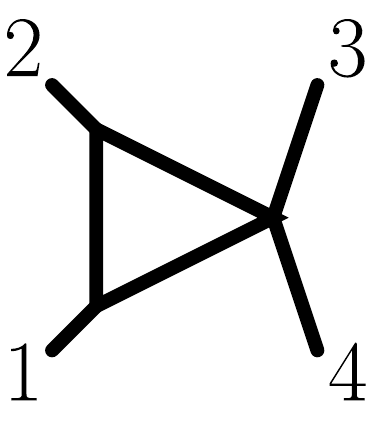}
 + \includegraphics[scale=0.5,trim= 0 60 0 0]{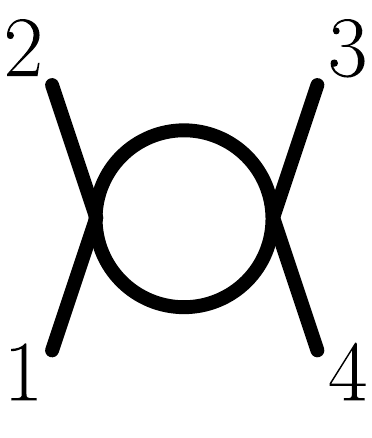}\,,\\ \nonumber
\end{align}
where $\omega$ is not relevant for the discussion.  The internal
propagators are all massless.  The triangle integral has only an
infrared divergence with a $1/\eps^2$ pole and the bubble has only an
ultraviolet divergence with a $1/\eps$ pole.  The $\eps$ dependence in
the coefficient of the triangle allows the infrared and ultraviolet
divergences to mix.  In order to directly extract ultraviolet
divergences without introducing an explicit infrared cutoff (such as a
mass) we must make sure that the IBPs being used do not mix infrared
and ultraviolet poles.  These subtleties are pertinent to our
discussion since our aim is to extract ultraviolet divergences by
focusing on scaleless vacuum integrals, which vanish in dimensional
regularization.

However, IBP identities that avoid both of the above complications can
be directly used to give relations between the ultraviolet divergences
of different dimensionally-regularized vacuum integrals without
introducing an additional explicit infrared cutoff.  In this way we
can demonstrate ultraviolet cancellations without explicitly
evaluating any integrals.  The situation in the presence of
subdivergences is more subtle and outside the scope of our present
discussion. We note that our principal aim is to examine the loop
order where ultraviolet divergences might first occur, so
subdivergences are not of primary concern.

Consider the following identities between two-loop vacuum integrals
\begin{align}
\text{UV finite} &= \int d^D p \, d^D q \,
\left( p^\mu \frac{\partial}{\partial p^\mu}
- q^\mu \frac{\partial}{\partial q^\mu} \right)
\frac{1}{ (p^2)^A (q^2)^B ((p+q)^2)^C} \nonumber \\
&= (-2A+2B) \, I_{A,B,C} - 2C \, I_{A-1, B, C+1} + 2C \, I_{A,B-1,C+1} \,,\nonumber \\
\text{UV finite} &= \int d^D p \, d^D q \,
\left( p^\mu \frac{\partial}{\partial q^\mu} \right)
\frac{1}{ (p^2)^A (q^2)^B ((p+q)^2)^C} \nonumber \\
&= (-B+C) \, I_{A,B,C} - B \, I_{A-1, B+1, C} + B \, I_{A, B+1, C-1} \nonumber \\
&\quad + C \, I_{A-1,B,C+1} - C \, I_{A,B-1,C+1}\,, \nonumber \\
\text{UV finite} &= \int d^D p \, d^D q \,
\left( q^\mu \frac{\partial}{\partial p^\mu} \right)
\frac{1}{ (p^2)^A (q^2)^B ((p+q)^2)^C} \nonumber \\
&= (-A+C) \, I_{A,B,C} - A \, I_{A+1, B-1, C} + A \, I_{A+1, B, C-1} \nonumber \\
&\quad + C \, I_{A,B-1,C+1} - C \, I_{A-1,B,C+1} \,.
\label{eq:logIBP5d}
\end{align}
In any of the three above identities, we can easily write the
integrand as a total derivative because the contributions arising from
commuting the loop momenta past the derivatives vanish. As desired
there is no explicit dependence on the dimension $D$.  With $A+B+C=5$,
the above IBP identities relate logarithmically divergent integrals in
$D=5$.

With dimensional regularization (and a mass as infrared cutoff) there
are no boundary terms, but here we allow more general regularization
schemes, in which case there may be a ultraviolet finite boundary term
on the left hand side of Eqs.\ \eqref{eq:logIBP5d}.  As elaborated in
the appendix, even in such schemes, boundary terms do not contain
divergences and do not modify the relations.  We therefore use
\eqn{eq:logIBP5d} as a direct relationship between the ultraviolet
divergences of the vacuum integrals.

With $A=1,B=C=2$, the first equation in Eqs.\ \eqref{eq:logIBP5d} provides the following
relation between the leading overall divergences of the integrals
\begin{align}
\left( I_{1,2,2} + 2 I_{1,1,3} - 2 I_{0,2,3} \right)\big |_{\rm log\ UV}
=  \left( I_{1,2,2} + 2 I_{1,1,3} \right)\big |_{\rm log\ UV} =0 \,,
\label{eq:keyIBP}
\end{align}
where we used the fact that $I_{0,2,3}$ is a ``one-loop squared''
integral with power divergences and no logarithmic divergence. This is
consistent with the explicit results in \eqn{eq:divs},
while allowing us to expose cancellations in \eqn{eq:PandNPdiv}
without computing divergences of individual integrals or using
identities that depend on details of the integrand.

In addition, by starting with the Yang--Mills integrand from
Ref.~\cite{DoubleCopyUnitarity} to construct the half-maximal
supergravity integrand via \eqn{TwoLoopDoubleCopy}, we have checked
that for {\it any} external state, the log divergences in $D=5$ are
always proportional to the same combination as above,
\begin{equation}
(I_{1,2,2} + 2I_{3,1,1})\,,
\end{equation}
whose leading log divergence vanishes.

While dimensional regularization is not sensitive to the potential
quadratic divergences in $D=5$, we can study these divergences by
lowering the dimension to $D=4$.  In $D=4$ one finds that for any
helicity configuration $h$ the expanded amplitude is
\begin{equation}
\mathcal A_h = C_h \left( 2 I_{3,3,-2} - 11 I_{3,2,-1}
+ 7 I_{3,1,0} +5 I_{2,2,0} \right)
 + \text{UV finite}\,,
\label{eq:4dvacResult}
\end{equation}
for some coefficient $C_h$ depending on the external states and on
choices made for reference momenta when choosing external
polarizations.  We constructed the required integrand by starting from
two-loop four-point Feynman diagrams for pure-Yang-Mills and then
applied to double-copy procedure to generate the diagrams
of half-maximal supergravity.  These are then expanded
large loop momentum and simplified using Lorentz symmetry
to obtained  \eqn{eq:4dvacResult}.
We apply the identities \eqref{eq:logIBP5d} to
the $D=4$ case, under the logarithmic power-counting requirement
$A+B+C=4$, with $A,B,C$ chosen to be all possible combinations of
integers (some of which may be negative) with some cutoff on their
absolute values. Dozens of IBP identities are generated, and the
resulting linear system relates all integrals to $I_{1,2,2}$. In this
way, we obtain cancellation of the divergences of \eqn{eq:4dvacResult}
for the vacuum expansion of the $\mathcal N=4$ supergravity amplitude.

Thus, we see that the two-loop cancellations in $D=4$ and $D=5$ can be
understood entirely and systematically using IBP identities.


\subsection{Generalizations and an all-loop conjecture}

In general, the structure of IBP equations can be rather opaque.
Might there be a simple organizing principle that applies to all loop
orders?  A strong hint is that the subset of IBP identities given in
\eqn{Lorentz} follows from Lorentz symmetry. We also saw the key role
that Lorentz symmetry played at one loop in \sect{sec:example}.  The
obvious $L$-loop extension is
\begin{equation}
\text{UV finite } = \int \biggl( \prod_{a=1}^L d^D \ell_a \biggr) \, v_i^{[\mu} v_j^{\nu]}
\sum_{a=1}^L \ell_{a\mu} \frac{\partial}{\partial \ell_a^\nu}
\frac{\mathcal N(\ell_a\cdot v_b, \ell_a \cdot \ell_b)}
{ \prod_j D_j^{A_j} } \,,
\label{LorentzL}
\end{equation}
where the $\ell_a$ are an independent set of loop momenta to be
integrated, the $v_a$ a set of external vectors in the problem and the
$1/D_j$ the propagators in the diagram.  As noted
earlier, we can equivalently apply Lorentz invariance following the
methods in Refs.~\cite{Simplifying,Mastrolia}.

What about the identities in \eqn{eq:logIBP5d}?  These can be
understood as belonging to a special class of IBP identities generated
by SL(2) transformations of the loop momenta of the form
\begin{equation}
 \begin{pmatrix}
  p \\
  q
 \end{pmatrix}
\rightarrow e^\omega
 \begin{pmatrix}
  p \\
  q
 \end{pmatrix},
\end{equation}
with some traceless $2 \times 2$ matrix $\omega$. Since such an
SL(2) transformation leaves the
integration measure $d^D p\, d^D q$ invariant, we have
\begin{equation}
\text{UV finite} = \int d^D p \, d^D q\, \omega_{ab}\, \ell_a^\mu \frac{\partial}{\partial \ell_b^\mu}
   \frac{1}{ {(p^2)}^{A} {(q^2)}^{B} {[(p+q)^2]}^{C} } \,,
\label{eq:sl2}
\end{equation}
where we used the notation $(\ell_1,\ell_2)=(p,q)$.  We can rewrite this as an IBP
relation,
\begin{equation}
\text{UV finite} = \int d^D p \, d^D q\, \frac{\partial}{\partial \ell_b^\mu}
   \frac{\omega_{ab}\,\ell_a^\mu}{ {(p^2)}^{A} {(q^2)}^{B} {[(p+q)^2]}^{C} }\, ,
   \label{eq:nodIBP}
\end{equation}
due to $\omega_{ab}$ being traceless. This also shows that these
relations do not have explicit dependence on the spacetime dimension
$D$.

In particular, the IBP identity which come from the first equation in
(\ref{eq:logIBP5d}) used to exhibit the cancellation of the logarithmic
divergence in $D=5$ is given by the SL(2) generator,
\begin{equation}
\omega_{ab} =
\begin{pmatrix}
  1 & 0 \\
  0 & -1
 \end{pmatrix}\,.
\end{equation}

\begin{figure}
\begin{align*}
\text{UV finite}  =
  \frac{1}{2} \includegraphics[scale=0.4,trim= -10 59 -10 0]{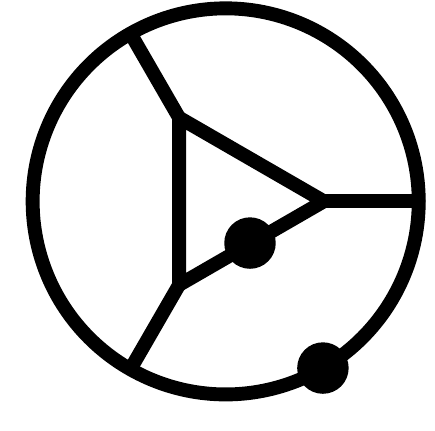}
          + 2 \includegraphics[scale=0.4,trim= -10 59 -10 0]{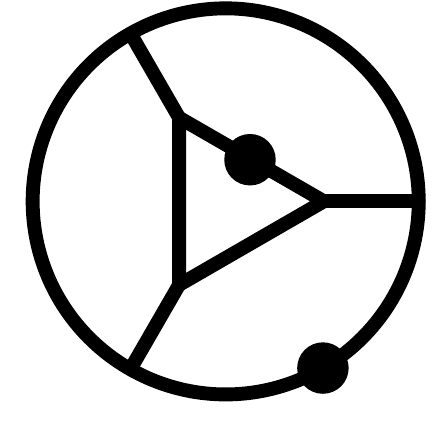}
          -   \includegraphics[scale=0.4,trim= -9 59 -10 0]{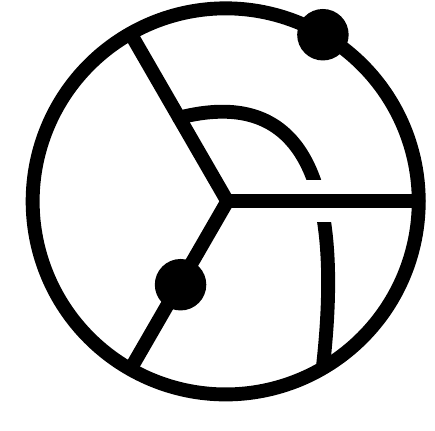}
          -   \includegraphics[scale=0.4,trim= -9 59 -10 0]{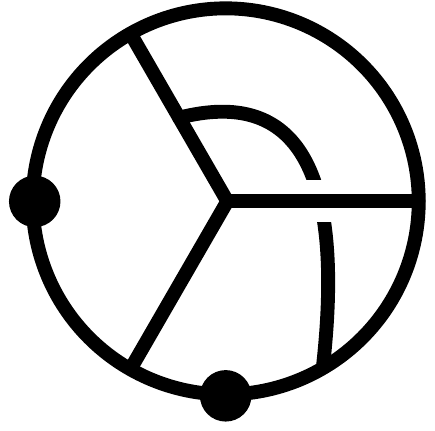} \\
\end{align*}
\caption{A four-loop relation between ultraviolet divergences of
  vacuum integrals in $D=11/2$ dimensions, matching identity 22 from
  Table I in Appendix C of Ref.~\cite{Simplifying}.  Where a black dot
  appears, the propagator is raised to a squared power.}
\label{fig:vacIdentity}
\end{figure}

In fact, the above ideas generalize trivially to the $L$-loop case by
considering generators of SL($L$). In more generality, the combination
of Lorentz invariance and SL($L$) transformations gives rise to some
subset of SL($D L$) transformations.  As a nontrivial check that these
ideas provide the key relations between the ultraviolet divergences of
vacuum integrals, we have reproduced the relations between ultraviolet
divergences of four-loop vacuum integrals in Appendix~C of
Ref.~\cite{Simplifying} in the context of obtaining the four-loop
ultraviolet divergence for $\NeqEight$ supergravity in the critical
dimension, $D=11/2$. One example of such a relation is given
graphically in \fig{fig:vacIdentity}.  This shows that Lorentz and
SL(4) symmetry generates a complete set of IBP identities necessary
for reducing the vacuum integrals encoding the ultraviolet divergence
to an independent set.  (We know the set is independent from
Eq.~(4.15) of Ref.~\cite{Simplifying}.)  In this case there were no
enhanced cancellations, but had they been present they would have been
found after applying the identities.

This brings us to a conjecture:
\begin{itemize}
\item
Given a loop integrand, homogeneous linear transformations of the loop
momentum variables with unit Jacobian are sufficient for revealing
enhanced cancellations of potential ultraviolet divergences in gravity
theories.
\end{itemize}
Generally, we are interested in the first divergence
of a theory in a given dimension so we do not need to concern
ourselves with complications due to subdivergences or divergences
beyond the logarithmic ones.  Even if the cancellation are not
complete and an ultraviolet divergence remains we expect these
symmetries to generate a complete set of IBP identities for
studying logarithmic divergences.

If this conjecture were to hold in general, it would shed light on the
mysterious enhanced cancellations that have been observed in various
supergravity theories.  Furthermore, these transformations can be
connected to the labeling difficulty of nonplanar
integrands. Remarkably, even though there does not seem to be a single
``discrete'' relabeling of the integration variables for each diagram
that allows us to construct an integrand that would manifest the
cancellations, the freedom to change integration variables appears to be
at the root of the cancellations.

\section{Conclusions}
\label{sec:conc}

In this paper we took initial steps towards systematically
understanding enhanced ultraviolet cancellations in supergravity
theories~\cite{N4GravThreeLoops, N5GravFourLoops, TwoLoopHalfMaxD5}.
These cancellations go beyond those presently understood from
standard-symmetry argumentation~\cite{VanishingVolume, KellyAttempt,
  HalfMaxMatter} and therefore appear to require novel explanations.

While a different avenue for understanding enhanced cancellations
based on exploiting the double-copy structure of gravity theories has
been successful for the special case of half-maximal supergravity in
$D=5$~\cite{TwoLoopHalfMaxD5}, it is unclear how to extend that
argument beyond two loops.  In contrast, our large loop-momentum
analysis here relies only on generic properties of the integrands and
integrals.

In nonabelian gauge theories, standard methods including superspace
techniques can be used expose ultraviolet cancellations at the
integrand level.  One might have thought that it is possible to
similarly find organizations of multi-loop integrands of supergravity
theory.  However, as we showed via one- and two-loop examples, it does
not seem possible to do this without relying also on integration
properties.

The simplest example of an enhanced cancellation in a supergravity
theory is probably the vanishing of one-loop divergences in pure
$\NeqFour$ supergravity in four dimensions. While the cancellation of
the divergence in $D=4$ is well understood as a consequence of
supersymmetry~\cite{OneLoopSugraDiv}, the pattern of cancellation
amongst the diagrams serves as a prototype for enhanced cancellations.
The double-copy construction~\cite{BCJLoop} allowed us to obtain the
$\NeqFour$ supergravity integrand very easily from the corresponding
ones of pure-Yang--Mills and $\NeqFour$ super-Yang--Mills theory.
Even in this relatively simple case where there are no labeling
ambiguities, we found that the cancellations cannot be exposed at
purely the integrand level.  After using integral identities that
follow from Lorentz invariance, the cancellations become visible.

We also investigated the more interesting case of half-maximal
supergravity at two loops.  In $D=5$, no standard symmetry explanation
is known for the cancellation that removes the logarithmic
divergence~\cite{VanishingVolume,TwoLoopHalfMaxD5}.  We showed that the
three-particle cuts display no integrand-level cancellations, even
though the final integrated expression does display the cancellations.
Based on our considerations, purely integrand-based proofs of the
observed enhanced cancellations do not appear to be possible.

In order to systematize ultraviolet cancellations after integration,
we used integration-by-parts identities~\cite{IBP}. This gives a
systematic means for finding all relations between the different
integrals.  While the machinery of doing so is generally difficult to
apply at high loop orders, at two-loops we made use of various
advances for controlling the complexity of the
identities~\cite{KosowerIBP, ItaIBP, LarsenZhang}.  As an
example we showed that one can use these ideas to rearrange the full
integrands of amplitudes so that they consist of terms that are
manifestly finite as well as terms that integrate to zero.  While this
construction is a proof of principle and gives some insight into how
the cancellations happen, it is too dependent on details of the
integrands and the associated identities to be useful for developing
an all-orders understanding.

To develop such an understanding, we instead focused on the large
loop-momentum behavior of the integrands.  For the two-loop $\NeqFour$
supergravity amplitude, by series expanding at large loop momentum, we
demonstrated that the only identities needed to expose the
cancellation are those that follow from Lorentz and an SL(2) symmetry.
Using these principles we also reproduced the necessary four-loop
identities~\cite{Simplifying} for extracting the ultraviolet
divergence of $\NeqEight$ in the critical dimension where it first
appears, suggesting that we have identified the key identities.

This led us to conjecture that at $L$ loop order the integral
identities generated by Lorentz and SL$(L)$ symmetry are sufficient
for exposing the enhanced cancellations of ultraviolet divergences,
when they happen.  If generally true, it would point towards a
symmetry explanation of enhanced cancellations.

There are a number of avenues for further exploration. It would be
important to first explicitly confirm our conjecture for the known
three- and four-loop examples of enhanced ultraviolet
cancellations~\cite{N4GravThreeLoops,N5GravFourLoops}, and to develop
an all-loop understanding.  It would also be interesting to study
whether this set of integral identities is also applicable to more
general problems in QCD and other theories that involve extracting
ultraviolet divergences.  It may also turn out to be helpful for
efficiently obtaining the required integration-by-parts identities
for analyzing divergences in $\NeqEight$ supergravity at five loops and beyond, once the integrands become available~\cite{DoubleCopy}.

We expect that in the coming years, as new theoretical tools are
developed, a complete and satisfactory understanding of enhanced
ultraviolet cancellations in gravity theories will follow.

\subsection*{Acknowledgments}

We thank Lance Dixon, Enrico Hermann, Harald Ita, David Kosower, James
Stankowicz, Jaroslav Trnka, and Yang Zhang for many enlightening
discussions.  This material is based upon work supported by the
Department of Energy under Award Number DE-{S}C0009937.  J.P.-M. is
supported by the U.S.~Department of State through a Fulbright
Scholarship.

\appendix

\section{Boundary terms in logarithmically divergent IBPs}
\label{sec:logIBP}

In section~\ref{sec:ibp} we claimed that for logarithmically divergent
integrals even in schemes other than dimensional regularization, the
boundary contributions of the IBP relations do not alter the relation
between the divergences.  Here we demonstrate this.
This is relevant to our discussion because it supports
the notion that the required IBP relations to obtain
the cancellations of the studied logarithmic divergences
are robust and do not depend on details of the scheme.

First, recall that the vacuum expansion to logarithmically
divergent integrals, the IBPs are of the form,
\begin{equation}
\int \prod_i d^D \ell_i \, \frac{\partial}{\partial \ell_j^\mu}
\left( \frac{\ell_k^\mu \prod_a N_a^{B_a}} {\prod_b D_b^{A_b}}\right) \,,
\label{eq:logIBP}
\end{equation}
where the powers $A_b$ and $B_a$ of the propagators $1/D_b$ and
irreducible numerators $N_a$ are such that the integrals are
logarithmically divergent.  Consider ultraviolet regularization after
Wick rotation using a physical cut off $\Lambda$, under which the
right-hand-side of \eqn{eq:logIBP}, as a total divergence, is turned
into a boundary integral at the compact cutoff surface by Stokes'
theorem. Since the number of propagators makes the integral
logarithmically divergent, the boundary integral also has mass
dimension $0$.  In Wilson's
floating cutoff picture, a change in the cutoff $\Lambda$ does not change the
boundary integral, which precludes it from having an ultraviolet
divergence. Note that the above argument breaks down if we consider, e.g.
quadratically divergent IBP relations. This argument is equivalent to
the textbook explanation of the finiteness of anomalies in one-loop
diagrams given by a boundary term of a linearly divergent
integral~\cite{AnomalyTextBook}.

However, there is an extra subtlety at higher loops that does not
arise in the study of anomalies.  The argument cannot be trivially
extended to the case where there are subdivergences because there is
no longer just one UV divergence coefficient to be fixed by a single floating cutoff. However, this is of secondary
concern because usually we are interested in studying the very first
potential divergence of a supergravity theory.  (There are some
subtleties with evanescent effects feeding into divergences which
require some care~\cite{TwoLoopEvanescent}.)  The most interesting
cases, such as $\NeqEight$ supergravity at five loops in $D=24/5$,
automatically have no subdivergences because of a lack of lower-loop
divergences.  It would be nevertheless interesting to understand the
behavior of boundary terms in general and study whether the relations
generated by Lorentz and SL($L$) symmetry can be applied to more
general problems of extracting divergences from vacuum integrals in
the presence of subdivergences.

We also comment on the dimensional regularization, which requires a
mass regulator to separate out infrared singularities. One might worry
that this mass regulator might interfere with the IBP identities.
However, it is easy to argue that when there are no subdivergences the
mass regulator does not cause any issues.
To prevent IBP identities from mixing up ultraviolet
and infrared poles, infrared divergences can be regulated by introducing a uniform
mass $m$ to every propagator on the right-hand-side of
\eqn{eq:logIBP}. It is best to introduce the mass prior to vacuum
expansion to retain cancellations of subdivergences~\cite{Vladimirov}.
After series expanding in small external momentum, we again obtain a
sum of logarithmically divergent vacuum integrals (whose internal
propagators are regulated by the uniform mass), but we also obtain
additional vacuum integrals multiplied by factors of $m^2$. To have
the correct dimensions, these additional integrals must have negative
mass dimension and are power-counting finite in the ultraviolet.
Assuming there are no one-loop subdivergences, a naive power counting
is sufficient for establishing the lack of ultraviolet
divergence. Therefore we obtain relations between logarithmic
ultraviolet divergences of massive vacuum integrals. Furthermore,
there is a smooth limit when the dimension $D$ tends to a fixed
integer (or a fractional number in more exotic cases), while the mass
$m$ tends to zero, because our special IBP identities have no $D$
dependence and because leading logarithmic ultraviolet divergences are
mass-independent. So we end up with relations between logarithmic
ultraviolet divergences of massless vacuum integrals.  This argument is applicable whenever
dimensional regularization rules out lower-loop subdivergences, for
example for supergravity calculations in fractional dimensions (see
e.g., Ref.~\cite{Simplifying}). We note that
Ref.~\cite{SimonJohannesFiniteInts} also investigated well-defined
limits of IBP identities as the dimension tends to an integer, in the
different context of studying finite integrals.


\end{document}